\documentclass[12pt]{article}
\usepackage{a4wide,amsmath,graphicx}
\numberwithin{equation}{section}
\DeclareMathOperator{\Tr}{Tr}
\newcommand{\QQ}{\bar{Q}_v}
\newcommand{\Q}{Q^*_v}
\newcommand{\Qq}{Q_v}
\newcommand{\DD}{\rlap{\hspace{0.2em}/}D}
\newcommand{\Li}[1]{\mathop{\mathrm{Li}}\nolimits_{#1}}

\title{$B$-meson distribution amplitudes%
\thanks{Lectures given at the Int.\ School on Heavy Quark Physics,
Dubna, June 2005}}
\author{Andrey Grozin\\
Budker Institute of Nuclear Physics}
\date{}

\begin{document}

\maketitle

\begin{abstract}
$B$-meson light-cone distribution amplitudes
are discussed in these lectures in the framework of HQET.
The evolution equation for the leading-twist distribution amplitude
is derived in one-loop approximation.
QCD sum rules for distribution amplitudes are discussed.
\end{abstract}

\section{Introduction}
\label{S:Intro}

Many exclusive $B$-decay amplitudes in the framework of SCET
contain $B$-meson distribution amplitudes~\cite{BF:01}.
The amplitude of the decay $B\to\gamma l\bar{\nu}$ at large photon energies
is given, up to power corrections, by a convolution of a hard part
(perturbatively calculable)
and the $B$-meson distribution amplitude~\cite{BHLN:03}.
Amplitudes of some decays, e.g.,
$B\to\pi l\bar{\nu}$ at large pion energies,
contain both factorizable
and non-factorizable contributions~\cite{BF:04,LN:04}.
Factorizable parts of decay amplitudes contain
light-cone distribution amplitudes of the initial $B$-meson
and final hadron(s).
They describe large-distance (soft) structure of these hadrons,
and cannot be calculated in perturbation theory.
The theory of hadronic distribution amplitudes in QCD
is reviewed in~\cite{CZ:84}.

Quark--antiquark distribution amplitudes of $B$-meson in HQET
were introduced and investigated in~\cite{GN:97}.
They are defined as Fourier transforms of matrix elements
of some gauge-invariant bilocal operators
between $B$-meson and vacuum.
Renormalization of these operators was calculated
in one-loop approximation.
However, an unusual term $1/\varepsilon^2$
in the one-loop renormalization constants
was erroneously omitted in~\cite{GN:97}.
The correct evolution equation
for the leading-twist distribution amplitude
was derived in~\cite{LN:03} at one loop.
The evolution kernel contains,
in addition to terms obtained earlier~\cite{GN:97},
an unusual term $\log(\omega/\mu)\delta(\omega-\omega')$.
The method of solution of the evolution equation
is also discussed in~\cite{LN:03} in detail.

Quark--antiquark--gluon distribution amplitudes of $B$-meson
and their relations to quark--antiquark ones
(based on equations of motion) are discussed in~\cite{KKQT:01}.

Sum rules for the quark--antiquark distribution amplitudes
were obtained in~\cite{GN:97}.
A simple model of these distribution amplitudes
at a low normalization scale (of order of hadronic scale)
was proposed.
Radiative corrections to the perturbative term
and the quark-condensate term were later calculated~\cite{BIK:04}.

In these lectures, we first briefly discuss
what is HQET%
\footnote{Unlike most texts on HQET,
we consider not a heavy quark but a heavy antiquark.
Of course, this makes no difference,
but the active participant in the distribution amplitudes
is the light quark (and, possibly, a gluon etc.),
and this choice makes notation more natural.}
(Sect.~\ref{S:HQET}).
A much more detailed presentation can be found
in the textbooks~\cite{MW:00,G:04}.
After a short discussion of $f_B$ (Sect~\ref{S:fB}),
quark--antiquark distribution amplitudes
are introduced (Sect.~\ref{S:Qq}).
Quark--antiquark--gluon distribution amplitudes
and their relations to two-particle ones
are discussed in Sect.~\ref{S:Qqg}.
Sect.~\ref{S:Evol} is the central (and longest) one.
Here renormalization of light-cone bilocal quark operators
in HQET is considered in one-loop approximation.
A detailed derivation of the evolution kernel is presented,
based on the methods of~\cite{MR:85}.
Finally, sum rules for the distribution amplitudes
are briefly discussed in Sect.~\ref{S:SR}.

\section{Heavy Quark Effective Theory}
\label{S:HQET}

Let's consider a heavy antiquark with mass $m$ and momentum
\begin{equation}
p = mv + k
\label{ResMom}
\end{equation}
in QCD.
Here $v$ is a fixed 4-velocity ($v^2=1$),
and the residual momentum $k$,
as well as momenta of all light quarks and gluons,
are supposed to be small (compared to $m$).
The propagator of this heavy antiquark can be written as
\begin{equation}
\frac{m - m \rlap/v - \rlap/k}{(mv+k)^2 - m^2 + i0}
= \frac{1-\rlap/v}{2}\,\frac{1}{k\cdot v + i0}
+ \mathcal{O}\left(\frac{1}{m}\right)\,.
\label{QProp}
\end{equation}
The leading term here is the HQET propagator.
This can be graphically presented as%
\footnote{The arrow here is somewhat misleading:
there is a \emph{particle} called antiquark and propagating from left to right;
it has no antiparticle.}
\begin{equation}
\raisebox{-5.5mm}{\begin{picture}(22,8)
\put(11,5){\makebox(0,0){\includegraphics{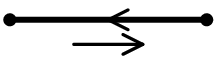}}}
\put(11,0){\makebox(0,0)[b]{$mv+k$}}
\end{picture}}
= \raisebox{-5.5mm}{\begin{picture}(22,8)
\put(11,5){\makebox(0,0){\includegraphics{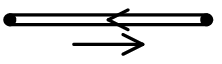}}}
\put(11,0){\makebox(0,0)[b]{$mv+k$}}
\end{picture}}
+ \mathcal{O}\left(\frac{1}{m}\right)\,.
\label{GraphProp}
\end{equation}
In a vertex sandwiched between such propagators,
we may substitute
\begin{equation}
\frac{1-\rlap/v}{2} \gamma^\mu \frac{1-\rlap/v}{2}
= \frac{1-\rlap/v}{2} (-v^\mu) \frac{1-\rlap/v}{2}\,.
\label{VertexSub}
\end{equation}
Projector can also be inserted near external legs,
so that all QCD vertices can be replaced by the HQET ones,
\begin{equation}
\raisebox{0mm}{\begin{picture}(12,9)
\put(6,3.5){\makebox(0,0){\includegraphics{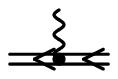}}}
\put(4,9){\makebox(0,0)[t]{$\mu$}}
\put(8,9){\makebox(0,0)[t]{$a$}}
\end{picture}}
= i g_0 t^a (-v^\mu)\,,
\label{Vertt}
\end{equation}
up to $\mathcal{O}(k/m)$ corrections.

These Feynman rules can be obtained from the Lagrangian
\begin{equation}
L = \QQ i v \cdot \overleftarrow{D} Q_v
+ (\text{light fields})\,
\label{HQETlagr}
\end{equation}
The antiquark field is our main field here, it satisfies
\begin{equation}
\QQ \rlap/v = - \QQ\,;
\label{LowerCompon}
\end{equation}
$Q_v$ is the conjugate field for $\QQ$.
Here the covariant derivatives are
\begin{equation}
D_\mu q = \left( \partial_\mu - i A_\mu \right) q\,,\quad
\bar{q} \overleftarrow{D}_\mu
= \bar{q} \left( \overleftarrow{\partial}_\mu + i A_\mu \right)\,,\quad
A_\mu = g_0 A^a_{0\mu} t^a\,.
\label{CovarDer}
\end{equation}
QCD tree diagrams are reproduced by HQET
up to $\mathcal{O}(k_i/m)$ corrections
(they can also be reproduced,
if we add the appropriate $1/m$ terms into the Lagrangian).

The heavy quark chromomagnetic moment is $\sim1/m$ by dimensionality.
At the leading order in $1/m$, the heavy-quark spin
does not interact with the gluon field.
Therefore, it may be rotated at will, without changing the physics
(heavy-quark spin symmetry).
It may even be switched off (superflavour symmetry).
We shall work with the spinless heavy antiquark,
\begin{equation}
L = \Q i v \cdot \overleftarrow{D} \Qq
+ (\text{light fields})\,,
\label{Spin0}
\end{equation}
during most of these lectures,
because this greatly simplifies reasoning and calculations.
Here again $\Q$ is the main (scalar) field,
and $\Qq$ is its conjugate.

\begin{figure}[ht]
\begin{center}
\begin{picture}(40,11.5)
\put(20,4){\makebox(0,0){\includegraphics{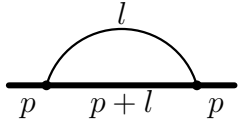}}}
\put(20,-4){\makebox(0,0)[b]{$p+l$}}
\put(20,11.5){\makebox(0,0)[t]{$l$}}
\put(7.5,-4){\makebox(0,0)[b]{$p$}}
\put(32.5,-4){\makebox(0,0)[b]{$p$}}
\end{picture}
\end{center}
\caption{Heavy--light two-point diagram}
\label{F:thresh}
\end{figure}

So, tree QCD diagrams, expanded in $k_i/m$ to some order,
are reproduced by the corresponding HQET diagrams.
But what about loops?
Here things are not so simple~\cite{S:01}.
Let's consider, for example, the heavy--light two-point diagram
(Fig.~\ref{F:thresh}) with $p=mv+k$,
where the residual momentum $k$ is small.
By choosing $v$ along $p$ we can always ensure $k=\omega v$.
Let's consider the integral
\begin{equation}
I = \frac{-i m^2}{\pi^{d/2}} \int
\frac{d^d l}{\left[m^2-(mv+k+l)^2\right]^2 (-l^2)}\,,
\label{ThresholdInt}
\end{equation}
which has neither ultraviolet (UV) nor infrared (IR) divergences.
There are two regions of the loop momentum $l$ in this integral:
\begin{itemize}
\item Hard region $l \sim m$.
Expanding the integrand in $k\ll m$, $l$, we have
\begin{equation}
\frac{1}{\left[m^2-(mv+l)^2\right]^2 (-l^2)}
+ 4 \frac{(m + l \cdot v) \omega}{\left[m^2-(mv+l)^2\right]^3 (-l^2)}
+ \cdots
\label{HardIntegrand}
\end{equation}
\item Soft region $l \sim \omega$.
Expanding the integrand in $k$, $l\ll m$, we have
\begin{equation}
\frac{1}{\left[-2 m (k+l) \cdot v\right]^2 (-l^2)}
+ 2 \frac{(k+l)^2}{\left[-2 m (k+l) \cdot v\right]^3 (-l^2)}
+ \cdots
\label{SoftIntegrand}
\end{equation}
\end{itemize}

\begin{figure}[ht]
\begin{center}
\begin{picture}(32,11.5)
\put(16,4){\makebox(0,0){\includegraphics{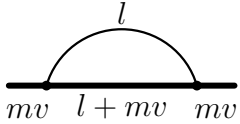}}}
\put(16,-4){\makebox(0,0)[b]{$l+mv$}}
\put(16,11.5){\makebox(0,0)[t]{$l$}}
\put(3.5,-4){\makebox(0,0)[b]{$mv$}}
\put(28.5,-4){\makebox(0,0)[b]{$mv$}}
\end{picture}
\end{center}
\caption{On-shell massive two-point integrals}
\label{F:OS1}
\end{figure}

The contribution of the hard region is
\begin{equation}
\begin{split}
&I_h = m^{-2\varepsilon} \left[ M(2,1)
+ 2 \frac{\omega}{m} \left[M(3,0) - M(2,1) + 2 M(3,1)\right]
+ \cdots \right]\,,\\
&\frac{\mu^{2\varepsilon} I_h}{\Gamma(1+\varepsilon)}
= - \frac{1}{2\varepsilon} + \log\frac{m}{\mu}
+ \left( \frac{1}{\varepsilon} - 2 \log\frac{m}{\mu} - 1 \right)
\frac{\omega}{m} + \cdots
\end{split}
\label{HardContr}
\end{equation}
where the on-shell massive two-point integrals (Fig.~\ref{F:OS1})
\begin{equation}
\begin{split}
&\int \frac{d^d l}{D_1^{n_1}D_2^{n_2}} = i \pi^{d/2} m^{d-2(n_1+n_2)} M(n_1,n_2)\,,\\
&D_1 = m^2 - (l+mv)^2 - i0\,,\quad
D_2 = -l^2-i0
\end{split}
\label{Mdef}
\end{equation}
are
\begin{equation}
M(n_1,n_2) =
\frac{\Gamma(d-n_1-2n2)\Gamma(-d/2+n_1+n_2)}{\Gamma(n_1)\Gamma(d-n_1-n_2)}\,.
\label{Mres}
\end{equation}
This contribution is IR divergent.

\begin{figure}[ht]
\begin{center}
\begin{picture}(32,11.5)
\put(16,4){\makebox(0,0){\includegraphics{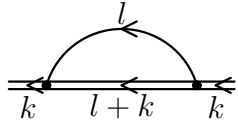}}}
\put(16,-4){\makebox(0,0)[b]{$l+k$}}
\put(16,11.5){\makebox(0,0)[t]{$l$}}
\put(3.5,-4){\makebox(0,0)[b]{$k$}}
\put(28.5,-4){\makebox(0,0)[b]{$k$}}
\end{picture}
\end{center}
\caption{HQET two-point integrals}
\label{F:HQET1}
\end{figure}

The contribution of the soft region is
\begin{equation}
\begin{split}
&I_s = (-2\omega)^{-2\varepsilon} \left[ I(2,1)
+ \frac{\omega}{m} \left[ I(3,1) - 2 I(2,1) \right] + \cdots \right]\,,\\
&\frac{\mu^{2\varepsilon} I_s}{\Gamma(1+\varepsilon)}
= \frac{1}{2\varepsilon} - \log\frac{-2\omega}{\mu}
- \left( \frac{1}{\varepsilon} - 2 \log\frac{-2\omega}{\mu}
- \frac{1}{2} \right) \frac{\omega}{m} + \cdots
\end{split}
\label{SoftContr}
\end{equation}
where the HQET two-point integrals (Fig.~\ref{F:HQET1}, $\omega = k \cdot v$)
\begin{equation}
\begin{split}
&\int \frac{d^d l}{D_1^{n_1}D_2^{n_2}} = i \pi^{d/2} (-2\omega)^{d-n_1-2n_2} I(n_1,n_2)\,,\\
&D_1 = - 2 (l+p) \cdot v - i0\,,\quad
D_2 = -l^2-i0
\end{split}
\label{Idef}
\end{equation}
are
\begin{equation}
I(n_1,n_2) = \frac{\Gamma(-d+n_1+2n_2)\Gamma(d/2-n_2)}{\Gamma(n_1)\Gamma(n_2)}\,.
\label{Ires}
\end{equation}
This contribution is UV divergent.

The complete result is finite:
\begin{equation}
I = - \log\frac{-2\omega}{m}
+ \left( 2 \log\frac{-2\omega}{m} - \frac{1}{2} \right)
\frac{\omega}{m} + \cdots
\end{equation}
This expansion can be easily continued if desired.

\begin{figure}[ht]
\begin{center}
\begin{picture}(152,18)
\put(16,9){\makebox(0,0){\includegraphics{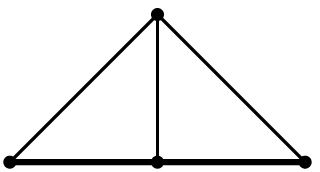}}}
\put(56,9){\makebox(0,0){\includegraphics{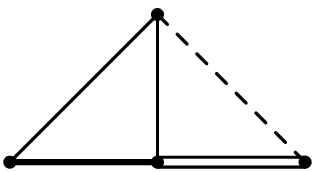}}}
\put(96,9){\makebox(0,0){\includegraphics{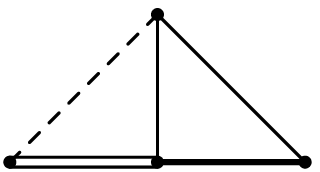}}}
\put(136,9){\makebox(0,0){\includegraphics{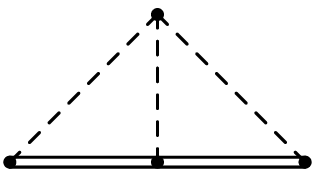}}}
\end{picture}
\end{center}
\caption{Regions in the two-loop diagram}
\label{F:Regions}
\end{figure}

What about higher loops?
Let's consider, for example, the two-loop two-point diagram
with a small external residual momentum.
There are several regions of the loop momenta in this diagram
(Fig.~\ref{F:Regions}).
In each of them, some momenta are hard ($\sim m$),
some are soft ($\sim\omega$).
As we have already seen,
a heavy-quark line with a soft residual momentum
becomes an HQET line.
Soft massless lines are shown by dashed lines in the figure.

For example, in the second diagram in Fig.~\ref{F:Regions},
the left loop is hard.
It contains a single scale $m$.
All external momenta of this loop are soft,
including those of the lines belonging to the other loop.
We can expand the integrand in these small momenta;
after taking the loop integral,
we obtain a polynomial in these momenta.
Now from the point of view of the soft loop
(large distances) this hard loop is just a local vertex.
This loop contains a single scale $\omega$,
and we can calculate it,
obtaining a non-analytical function
of the external residual energy $\omega$.

In a general multiloop diagram,
hard lines must always form loops,
so that momentum conservation can hold
after neglecting all soft momenta.
There can be several disconnected hard parts;
each of them must contain at least one heavy line
(a subdiagram consisting of only light lines and having
soft external momenta has no reason to be hard).
From the point of view of the soft part,
hard parts are just local vertices.
There may appear a soft subdiagram
connected to the rest of the diagram
only at such a vertex.
Such a subdiagram is scaleless and hence vanish.
For example, we could consider one more region
in Fig.~\ref{F:Regions}:
when the only soft line is the middle light line,
all the rest are hard.
But this soft line forms a loop containing one local vertex
(the integrated hard loop), and hence it vanishes.

In the usual HQET formalism,
Lagrangian contains local operators multiplied by matching coefficients.
QCD operators are also expanded in HQET operators
with matching coefficients.
These matching coefficients are the only quantities in the theory
which depend on the hard scale $m$.
Diagrammatically, they come from hard loops in QCD diagrams.
Local operators produce vertices polynomial in their external momenta.
They appear in HQET diagrams, which contain only the soft scale $\omega$.
These HQET diagrams are soft parts of QCD diagrams.

\section{$B$-meson decay constant}
\label{S:fB}

During these lectures, we shall mostly live in a world
with a heavy antiquark having $j^P=0^+$.
Physics in the real world is the same,
up to $1/m$ corrections.
We shall work in the $v$ rest frame.

The ground-state $S$-wave $\bar{Q} q$ meson
has the quantum numbers $j^P=\frac{1}{2}^+$.
There are 2 $P$-wave excited mesons
with $j^P=\frac{1}{2}^-$ and $\frac{3}{2}^-$.
The heavy--light quark current
\begin{equation}
j = \Q q
\label{j}
\end{equation}
has no definite parity;
the currents with parity $P=\pm1$ are
\begin{equation}
j_P = \frac{1+P\gamma^0}{2} j\,.
\label{jP}
\end{equation}
They have the quantum numbers of $S$-wave $\frac{1}{2}^+$ mesons
and $P$-wave $\frac{1}{2}^-$ mesons.
The ground-state meson $M$ has a Dirac wave function $u$
which satisfies $\gamma^0 u=u$ and is normalized by $\bar{u}u=1$.
The matrix element of $j$ from $M$ to vacuum is
\begin{equation}
{<}0|j|M{>} = F u\,,
\label{F}
\end{equation}
where the one-meson states are normalized by the non-relativistic condition
\begin{equation}
{<}M,\vec{p}\,'|M,\vec{p}\,{>} = (2\pi)^3 \delta(\vec{p}\,'-\vec{p}\,)\,.
\label{nrnorm}
\end{equation}

\begin{figure}[ht]
\begin{center}
\begin{picture}(56,24)
\put(28,14){\makebox(0,0){\includegraphics{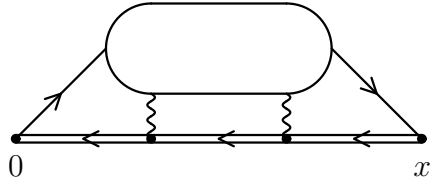}}}
\put(1,0){\makebox(0,0)[b]{$0$}}
\put(55,0){\makebox(0,0)[b]{$x$}}
\end{picture}
\end{center}
\caption{Correlator of heavy--light currents}
\label{F:cor}
\end{figure}

The correlator of the heavy--light currents (Fig.~\ref{F:cor}),
\begin{equation}
i {<}T j(x) \bar{\jmath}(0){>} = \delta(\vec{x}\,) \Pi(x^0)\,,
\label{cor1}
\end{equation}
has the structure
\begin{equation}
\Pi(x^0) = A + B \rlap/v\,.
\label{cor2}
\end{equation}
Therefore, the correlators of the currents with definite parity are
\begin{equation}
i {<}T j_P(x) \bar{\jmath}_P(0){>} = \delta(\vec{x}\,) \Pi_P(x^0)\,,\quad
\Pi_P = A + P B = \frac{1}{4} \Tr (1+P\gamma^0) \Pi\,.
\label{cor3}
\end{equation}
The spectral density of, say, the correlator with $P=+1$ is
\begin{equation}
\rho_+(\varepsilon) = F^2 \delta(\varepsilon-\bar{\Lambda}) + \cdots
\end{equation}
where $\bar{\Lambda}$ is the residual energy of the ground-state meson,
and the dots mean contribution of excited states.

Now we shall return to the real world
with $j^P=\frac{1}{2}^-$ heavy antiquark for a while,
but still with $m=\infty$.
The $S$-wave ground-state meson
turns into a degenerate doublet with $j^P=0^-$, $1^-$.
The two $P$-wave mesons turn into two degenerate doublets:
with $j^P=0^+$, $1^+$
and with $j^P=1^+$, $2^+$.
All heavy--light currents $\QQ\Gamma q$ reduce,
due to~(\ref{LowerCompon}), to 4 ones with
\begin{equation}
\Gamma = \gamma_5\,,\quad\vec{\gamma}
\qquad\text{and}\qquad
\Gamma = 1\,,\quad\vec{\gamma}\gamma_5\,.
\label{Gammai}
\end{equation}
The first 2 currents, with $\Gamma$ anticommuting with $\gamma^0$,
have the quantum numbers of the ground-state $0^-$, $1^-$ mesons;
the second 2 currents, with $\Gamma$ commuting with $\gamma^0$,
have the quantum numbers of the $P$-wave $0^+$, $1^+$ mesons.

The correlators (Fig.~\ref{F:cor}) are
\begin{equation}
i {<} T j_2(x) j_1^+(0) {>} = \delta(\vec{x}\,) \Pi_{12}(x^0)\,,\quad
\Pi_{12} = \Tr \overline{\Gamma}_1 \frac{1-\gamma^0}{2} \Gamma_2 \Pi\,,
\label{cor4}
\end{equation}
or
\begin{equation}
\Pi_{12} = \Pi_P \Tr \overline{\Gamma}_1 \frac{1-\gamma^0}{2} \Gamma_2\,,
\label{cor5}
\end{equation}
where $P=+1$ for $\Gamma$ anticommuting with $\gamma^0$
and $-1$ for commuting.
For $\Gamma=\gamma_5$ and $\gamma^i$,
the correlators are $2\Pi_+$ and $2\Pi_+\delta^{ij}$.
Their spectral densities are $F_B^2 \delta(\varepsilon-\bar{\Lambda})+\cdots$
and $F_{B^*}^2 \delta(\varepsilon-\bar{\Lambda}) \delta^{ij}+\cdots$,
where
\begin{equation}
{<}0|\QQ \gamma_5 q|B{>} = F_B\,,\quad
{<}0|\QQ \vec{\gamma} q|B^*{>} = F_{B^*} \vec{e}\,,
\label{FB}
\end{equation}
and dots mean contribution of higher states.
Therefore,
\begin{equation}
F_B = F_{B^*} = \sqrt{2} F\,.
\label{FB2}
\end{equation}

The usual definition of the decay constants is
\begin{equation}
{<}0|\QQ \gamma^\mu \gamma_5 q|B{>}_r = i f_B p^\mu\,,\quad
{<}0|\QQ \gamma^\mu q|B^*{>}_r = i m f_{B^*} e^\mu\,,
\label{fB}
\end{equation}
where the relativistic normalization
\begin{equation}
{}_r{<}B,p'|B,p{>}_r = (2\pi)^3 2 p^0 \delta(\vec{p}\,'-\vec{p}\,)
\label{relnorm}
\end{equation}
of single-meson states is used
(this normalization becomes meaningless in the limit $m\to\infty$,
and therefore cannot be used in HQET).
Comparing~(\ref{fB}) for $B$ or $B^*$ at rest with~(\ref{FB}),
we arrive at
\begin{equation}
f_B = f_{B^*} = \frac{2 F}{\sqrt{m}}\,,
\label{scaling}
\end{equation}
up to $1/m$ corrections.

Let's consider the correlator~(\ref{cor5}) with $P=+1$.
Equating the ground-state contributions to the spectral densities
of the left-hand side and the right-hand one, we obtain
\begin{equation}
{<}0|j_2|M{>} {<}M|j_1^+|0{>}
= F^2 \Tr \Gamma_2 \frac{1+\gamma^0}{2} \overline{\Gamma}_1\,.
\label{trf1}
\end{equation}
Therefore,
\begin{equation}
{<}0|\QQ \Gamma q|M{>} = \frac{F}{\sqrt{2}}
\Tr \Gamma \frac{1+\gamma^0}{2} \Gamma_M\,,
\label{trf2}
\end{equation}
where the matrix
\begin{equation}
\Gamma_M =
\begin{cases}
- i \gamma_5 & \text{for $B$}\\
i \rlap/e & \text{for $B^*$}
\end{cases}
\label{GammaM}
\end{equation}
is defined up to a phase factor.
We can re-write this result for the relativistic normalization
of the meson state, in covariant notation:
\begin{equation}
{<}0|\QQ \Gamma q|M{>}_r = \sqrt{m} F \Tr \Gamma \mathcal{M}\,,
\label{trf}
\end{equation}
where
\begin{equation}
\mathcal{M} = \frac{1+\rlap/v}{2}
\begin{cases}
- i \gamma_5 & \text{for $B$}\\
i \rlap/e & \text{for $B^*$}
\end{cases}
\label{MMatrix}
\end{equation}
(of course, one can re-define the phases of $|M{>}$
and hence of $\mathcal{M}$).

\section{Quark--antiquark distribution amplitudes}
\label{S:Qq}

After safely returning to the ideal world with a $0^+$ heavy antiquark,
we want to invent an operator which probes more details of the structure
of $B$-meson than the local current~(\ref{j}).
To this end, we consider a bilocal gauge-invariant operator
\begin{equation}
\tilde{O}(t) = \Q(0) [0,z] q(z)\,,
\label{Ot0}
\end{equation}
where
\begin{equation}
[x,y] = P \exp \left[ - i \int_x^y A_\mu(z) d z^\mu \right]\,,
\label{Exy}
\end{equation}
and
\begin{equation}
z^2 = 0\,,\quad
t = v \cdot z\,.
\label{LightCone}
\end{equation}
Its matrix element from the ground-state meson to vacuum
has 2 Dirac structures:
\begin{equation}
{<}0|\tilde{O}(t)|M{>} = F
\left[ \tilde{\varphi}_+(t)
+ \frac{\tilde{\varphi}_-(t) - \tilde{\varphi}_+(t)}{2 t} \rlap/z \right] u\,.
\label{phit}
\end{equation}
because $\rlap/v u = u$.

In what follows, we shall often use light-front components of vectors.
Let's introduce (in the $v$ rest frame) two light-like vectors
\begin{equation}
\begin{split}
&n_\pm^\mu = (1, \mp 1, \vec{0}\,)\,,\\
&n_+^2 = n_-^2 = 0\,,\quad
n_+ \cdot n_- = 2\,.
\end{split}
\label{npm}
\end{equation}
Light-front components of any vector $a$ are defined as
\begin{equation}
a_\pm = a \cdot n_\pm = a^0 \pm a^1\,.
\label{apm}
\end{equation}
We have
\begin{equation}
\begin{split}
&a^\mu = \frac{1}{2} \left( a_+ n_-^\mu + a_- n_+^\mu \right) + a_\bot^\mu\,,\\
&a \cdot b = \frac{1}{2} \left( a_+ b_- + a_- b_+ \right)
- \vec{a}_\bot \cdot \vec{b}_\bot\,.
\end{split}
\label{scapro}
\end{equation}
In particular,
\begin{equation}
v^\mu = \frac{1}{2} \left( n_+^\mu + n_-^\mu \right)\,,\quad
v_+ = v_- = 1\,,\quad
\vec{v}_\bot = \vec{0}\,.
\label{vpm}
\end{equation}
We shall also use light-front components of $\gamma^\mu$:
\begin{equation}
\gamma_\pm = \gamma \cdot n_\pm = \rlap/n_\pm\,.
\label{gammapm}
\end{equation}

The definition~(\ref{phit}) can be re-written as
\begin{equation}
{<}0|\tilde{O}(t)|M{>} = \frac{1}{2} F
\left[ \tilde{\varphi}_+(t) \gamma_- + \tilde{\varphi}_-(t) \gamma_+ \right] u\,.
\label{Ot}
\end{equation}
If we introduce the operators
\begin{equation}
\tilde{O}_\pm(t) = \gamma_\pm \tilde{O}(t)\,,
\label{Opmt}
\end{equation}
then
\begin{equation}
{<}0|\tilde{O}_\pm(t)|M{>} = F \tilde{\varphi}_\pm(t) \gamma_\pm u\,.
\label{phipmt}
\end{equation}
The $B$-meson distribution amplitudes are the Fourier transforms
of these functions:
\begin{equation}
\varphi_\pm(\omega) = \frac{1}{2\pi}
\int \tilde{\varphi}_\pm(t) e^{i \omega t} dt\,,\quad
\tilde{\varphi}_\pm(t)
= \int \varphi_\pm(\omega) e^{-i \omega t} d\omega\,.
\label{phiom}
\end{equation}
They are normalized by
\begin{equation}
\tilde{\varphi}_\pm(0) = \int_0^\infty \varphi_\pm(\omega) d\omega = 1\,.
\label{norm}
\end{equation}
The function $\varphi_+(\omega)$ is the leading-twist distribution amplitude,
and $\varphi_-(\omega)$ -- the subleading-twist one
(though there is no good definition of twist in HQET).
We can formally introduce the operators
\begin{equation}
\begin{split}
&O_\pm(\omega) = \frac{1}{2\pi}
\int \tilde{O}_\pm(t) e^{i \omega t} dt
= \Q(0) \gamma_\pm \delta(i D_+ - \omega) q(0)\,,\\
&\tilde{O}_\pm(t) = \int O_\pm(\omega) e^{-i \omega t} d\omega\,,
\end{split}
\label{Oom}
\end{equation}
then
\begin{equation}
{<}0|O_\pm(\omega)|M{>} = F \varphi_\pm(\omega) \gamma_\pm u\,.
\label{Opmom}
\end{equation}
The distribution amplitudes describe the distribution
in the light-front component $p_+$ of the light-quark momentum
in $B$-meson.

The expansion of the operators~(\ref{Opmt}) in $t$ reads
\begin{equation}
\begin{split}
&\tilde{O}_\pm(t) = \sum_{n=0}^\infty O_\pm^{(n)} \frac{(-it)^n}{n!}\,,\\
&O_\pm^{(n)} = \int O_\pm(\omega) \omega^n d\omega
= \Q \gamma_\pm (i D_+)^n q\,,
\end{split}
\label{On}
\end{equation}
or for matrix elements
\begin{equation}
\begin{split}
&\tilde{\varphi}_\pm(t) = \sum_{n=0}^\infty {<}\omega^n{>}_\pm \frac{(-it)^n}{n!}\,,\\
&{<}\omega^n{>}_\pm = \int_0^\infty \varphi_\pm(\omega) \omega^n d\omega\,,\\
&{<}0|O_\pm^{(n)}|M{>} = F {<}\omega^n{>}_\pm \gamma_\pm u\,.
\end{split}
\label{omn}
\end{equation}
We can also reconstruct $O_\pm(\omega)$ from $O_\pm^{(n)}$:
\begin{equation}
O_\pm(\omega) =
\int_{-i\infty}^{+i\infty} O_\pm^{(n)} \omega^{-n-1} \frac{dn}{2\pi i}\,,
\label{intdn}
\end{equation}
or for matrix elements
\begin{equation}
\varphi_\pm(\omega) =
\int_{-i\infty}^{+i\infty} {<}\omega^n{>}_\pm \omega^{-n-1} \frac{dn}{2\pi i}
\label{intdn2}
\end{equation}
(you can easily check this by substituting~(\ref{On}) into~(\ref{intdn});
integration in $dn$ yields $\delta(\log(\omega'/\omega))$).

The first moments ${<}\omega{>}_\pm$ can be found from the equations of motion.
The equation of motion for the heavy antiquark is
\begin{equation}
\Q\overleftarrow{D}_0 = 0\,,
\label{heavyEOM}
\end{equation}
and therefore we obtain
\begin{equation}
{<}0|\Q D_0 q|M{>} = {<}0|\partial_0(\Q q)|M{>} = -i F \bar{\Lambda} u\,.
\label{D0}
\end{equation}
The vector part has the structure
\begin{equation}
{<}0|\Q \vec{D} q|M{>} = a F \vec{\gamma}\,u\,.
\label{Dvec1}
\end{equation}
The equation of motion of the light quark is
\begin{equation}
{<}0|\Q \DD q|M{>} = 0\,,
\label{lightEOM}
\end{equation}
and using $\DD=D_0\gamma^0-\vec{D}\cdot\vec{\gamma}$ we obtain
\begin{equation}
{<}0|\Q \vec{D} q|M{>} = \frac{i}{3} F \bar{\Lambda} \vec{\gamma} u\,.
\label{Dvec}
\end{equation}
Finally, we arrive at
\begin{equation}
{<}\omega{>}_+ = \frac{4}{3} \bar{\Lambda}\,,\quad
{<}\omega{>}_- = \frac{2}{3} \bar{\Lambda}\,.
\label{mom1}
\end{equation}

Let's now consider the second moments.
They involve two new (nonperturbative) hadronic parameters:
\begin{equation}
{<}0|\Q \vec{E}\cdot\vec{\alpha} q|M{>} = - i F \lambda_E^2 u\,,\quad
{<}0|\Q \vec{H}\cdot\vec{\sigma} q|M{>} = - F \lambda_H^2 u\,,
\label{lambdaEH}
\end{equation}
where
\begin{equation}
\vec{E} = i [D_0,\vec{D}]\,,\quad
\vec{H} = i \vec{D}\times\vec{D}\,,\quad
\vec{\alpha} = \gamma^0\vec{\gamma}\,,\quad
\vec{\sigma} = - \vec{\gamma} \gamma_5 \gamma^0\,.
\end{equation}
Now we can calculate all matrix elements with 2 derivatives.
From~(\ref{D0}) and~(\ref{Dvec}),
using the heavy-antiquark equation of motion~(\ref{heavyEOM}),
we immediately find
\begin{equation}
{<}0|\Q D_0^2 q|M{>} = - F \bar{\Lambda}^2 u\,,\quad
{<}0|\Q D_0 \vec{D} q|M{>} = \frac{1}{3} F \bar{\Lambda}^2 \vec{\gamma} u\,.
\label{D00}
\end{equation}
Using the definition~(\ref{lambdaEH}) of $\lambda_E^2$,
we immediately find
\begin{equation}
{<}0|\Q \vec{D} D_0 q|M{>} = \frac{1}{3} F
\left(\bar{\Lambda}^2 + \lambda_E^2\right) \vec{\gamma} u\,.
\label{Di0}
\end{equation}
The second spatial derivatives have the structure
\begin{equation}
{<}0|\Q D^i D^j q|M{>}
= F \left(b\delta^{ij}-\frac{i}{6}\lambda_H^2\varepsilon^{ijk}\sigma^k\right)u\,,
\label{Dij1}
\end{equation}
where the second coefficient follows
from the definition~(\ref{lambdaEH}) of $\lambda_H^2$.
We find $b$ using the light-quark equation of motion~(\ref{lightEOM}):
\begin{equation}
{<}0|\Q D^i D^j q|M{>} = - \frac{1}{3} F
\left[ \left(\bar{\Lambda}^2+\lambda_E^2+\lambda_H^2\right) \delta^{ij}
+ \frac{i}{2} \lambda_H^2 \varepsilon^{ijk}\sigma^k \right] u\,.
\label{Dij}
\end{equation}
Finally, we arrive at
\begin{equation}
{<}\omega^2{>}_+ = 2 \bar{\Lambda}^2
+ \frac{2}{3} \lambda_E^2 + \frac{1}{3} \lambda_H^2\,,\quad
{<}\omega^2{>}_- = \frac{2}{3} \bar{\Lambda}^2 + \frac{1}{3} \lambda_H^2\,.
\label{mom2}
\end{equation}

What about $B$-meson distribution amplitudes
in the real world with a $\frac{1}{2}^-$ heavy antiquark?
It has 4  distribution amplitudes, as any pseudoscalar meson:
\begin{equation}
\begin{split}
&{<}0|\Q(0) [0,z] \gamma_5 q(z)|B{>}_r = - i f_B m \tilde{\varphi}_P\,,\\
&{<}0|\Q(0) [0,z] \gamma^\mu \gamma_5 q(z)|B{>}_r
= f_B \left[ i \tilde{\varphi}_{A1} p^\mu - m \tilde{\varphi}_{A2} z^\mu \right]\,,\\
&{<}0|\Q(0) [0,z] \sigma^{\mu\nu} \gamma_5 q(z)|B{>}_r
= i f_B \tilde{\varphi}_T (p^\mu z^\nu - p^\nu z^\mu)\,.
\end{split}
\label{QCDps}
\end{equation}
Similarly to~(\ref{trf}), we have
\begin{equation}
{<}0|\Q(0) [0,z] \Gamma q(z)|M{>}_r = F \Tr \Gamma
\left[\tilde{\varphi}_+ + \frac{\tilde{\varphi}_--\tilde{\varphi}_+}{2t} \rlap/z\right]
\mathcal{M}\,.
\label{trfphi}
\end{equation}
Therefore these 4 QCD distribution amplitudes
can be expressed via 2 HQET ones:
\begin{equation}
\begin{split}
&\tilde{\varphi}_P = \frac{\tilde{\varphi}_+(t)+\tilde{\varphi}_-(t)}{2}\,,\quad
\tilde{\varphi}_{A1} = \tilde{\varphi}_+(t)\,,\\
&\tilde{\varphi}_{A2} = \tilde{\varphi}_T = \frac{i}{2}
\frac{\tilde{\varphi}_+(t)-\tilde{\varphi}_-(t)}{t}\,.
\end{split}
\label{QCDps2}
\end{equation}
The QCD distribution amplitudes are usually considered as functions
of the longitudinal momentum fraction $x=\omega/m$.
It is usually assumed that $\varphi_{A1}(x)\sim x$ at $x\to0$,
and $\varphi_P(x)\to\text{const}$.
Therefore, we shall assume that $\varphi_+(\omega)\sim\omega$ at $\omega\to0$,
and $\varphi_-(\omega)\to\text{const}$.
The QCD distribution amplitudes are normalized as
\begin{equation}
\tilde{\varphi}_P(0) = \tilde{\varphi}_{A1}(0) = 1\,,\quad
\tilde{\varphi}_{A2}(0) = \tilde{\varphi}_T(0) = \frac{\bar{\Lambda}}{3}
\label{QCDnorm}
\end{equation}
(to derive the last formula, we used the $t$ expansion~(\ref{omn})
and the first moments~(\ref{mom1})).

The vector meson $B^*$ is described by 6 distribution amplitudes in QCD.
All 10 distribution amplitudes (4 for $B$ plus 6 for $B^*$)
are expressed via 2 HQET distribution amplitudes $\varphi_\pm(\omega)$.
This is a consequence of the heavy-quark spin symmetry.
We don't present formulae for $B^*$ here;
they can be found in~\cite{GN:97}.

\section{Quark--antiquark--gluon distribution amplitudes}
\label{S:Qqg}

Now we shall discuss relations of quark--antiquark distribution amplitudes
and quark--antiquark--gluon ones following from the equations of motion.
In order to apply them, we need to differentiate with respect to
the coordinates of the light quark and the heavy antiquark separately.
Therefore, we go slightly off the light cone.
The generalization of~(\ref{phit}) to an arbitrary $z^2$ is
\begin{equation}
{<}0|\Q(0) q(z)|M{>} = F
\left[ \tilde{\varphi}_+(t,z^2)
+ \frac{\tilde{\varphi}_-(t,z^2)-\tilde{\varphi}_+(t,z^2)}{2t}
\rlap/z \right] u\,,
\label{phiz}
\end{equation}
where $t = v \cdot z$,
and the fixed-point gauge $x_\mu A^\mu(x)=0$ is used
to simplify notation.

In order to apply the light-quark equation of motion~(\ref{lightEOM}),
we apply
\begin{equation*}
\displaystyle \gamma^\mu \frac{\partial}{\partial z^\mu}
\end{equation*}
to this definition.
The differentiation of the right-hand side is straightforward.
In the left-hand side we write
\begin{equation*}
{<}0|\Q(0) \gamma^\mu (\partial_\mu - i A_\mu(z) + i A_\mu(z)) q(z)|M{>}\,.
\end{equation*}
The first two terms yield zero.
In the fixed-point gauge
\begin{equation}
A_\mu(z) = \int_0^1 G_{\nu\mu}(uz) u z^\nu du\,,
\label{fpg}
\end{equation}
where $G_{\mu\nu}=gG^a_{\mu\nu}t^a$.

The matrix element of the operator containing $G_{\mu\nu}$ contains 4 structures:
\begin{equation}
\begin{split}
&{<}0|\Q(0) [0,uz] i G_{\nu\mu}(uz) z^\nu [uz,z] q(z)|M{>} ={}\\
& - F \left[
(v_\mu \rlap/z - t \gamma_\mu) (\tilde{\psi}_A - \tilde{\psi}_V)
+ i \sigma_{\mu\nu} z^\nu \tilde{\psi}_V
- z_\mu \tilde{\psi}_X
+ \frac{z_\mu}{t} \rlap/z \tilde{\psi}_Y
\right] u\,,
\end{split}
\label{psi}
\end{equation}
and is parametrized by 4 quark--antiquark--gluon distribution amplitudes
(note that this matrix elements vanishes when multiplied by $z^\mu$).

Equating the coefficients of $u$ and $\rlap/z u$,
we obtain two consequences of the light-quark equation of motion:
\begin{align}
&\tilde{\varphi}_-' + \frac{\tilde{\varphi}_- - \tilde{\varphi}_+}{t}
= 2 t \int_0^1 (\tilde{\psi}_A - \tilde{\psi}_V) u\,du\,,
\label{eom1}\\
&\tilde{\varphi}_+' - \tilde{\varphi}_-'
+ \frac{\tilde{\varphi}_- - \tilde{\varphi}_+}{t}
+ 4 t \frac{\partial\tilde{\varphi}_+}{\partial z^2}
= 2 t \int_0^1 (2 \tilde{\psi}_V + \tilde{\psi}_A + \tilde{\psi}_X)
u\,du\,.
\label{eom2}
\end{align}

Similarly, to use the heavy-antiquark equation of motion we apply
\begin{equation*}
\Q(0) v^\mu \frac{\partial}{\partial z^\mu} q(z)
= v^\mu \partial_\mu (\Q(0) q(z))
- v^\mu \Q(0)
(\overleftarrow{\partial}_\mu + i A_\mu(0) - i A_\mu(0)) q(z)\,.
\end{equation*}
The first two terms in the last operator yield zero.
Now we move the heavy antiquark;
the center of the fixed-point gauge must be constant,
and we have to use the gauge $(x-z)^\mu A_\mu(x)=0$:
\begin{equation}
A_\mu(0) = - \int_0^1 G_{\nu\mu}(uz) (1-u) z^\nu du\,.
\label{fpg2}
\end{equation}
We arrive at two consequences of the heavy-antiquark equation of motion:
\begin{align}
&\tilde{\varphi}_+'
+ \frac{\tilde{\varphi}_- - \tilde{\varphi}_+}{2t}
+ i \bar{\Lambda} \tilde{\varphi}_+
+ 2 t \frac{\partial\tilde{\varphi}_+}{\partial z^2}
= - t \int_0^1 (\tilde{\psi}_A + \tilde{\psi}_X) (1-u) du\,,
\label{eom3}\\
&\tilde{\varphi}_-' - \tilde{\varphi}_+'
+ \frac{\tilde{\varphi}_+ - \tilde{\varphi}_-}{t}
+ i \bar{\Lambda} (\tilde{\varphi}_- - \tilde{\varphi}_+)
+ 2 t \left( \frac{\partial\tilde{\varphi}_-}{\partial z^2}
  - \frac{\partial\tilde{\varphi}_+}{\partial z^2} \right)
= 2 t \int_0^1 (\tilde{\psi}_A + \tilde{\psi}_Y) (1-u) du\,.
\label{eom4}
\end{align}

The functions
\begin{equation*}
\left.\frac{\partial\tilde{\varphi}_\pm(t,z^2)}{\partial z^2}\right|_{z^2=0}
\end{equation*}
are some new (non-leading) quark--antiquark distribution amplitudes.
We are not interested in them here.
There are 2 equations involving only our familiar distribution amplitudes
$\tilde{\varphi}_\pm(t,0)$, namely, (\ref{eom1})
and a combination of~(\ref{eom2}) and~(\ref{eom3}):
\begin{equation}
\begin{split}
&\tilde{\varphi}_-' + \frac{\tilde{\varphi}_- - \tilde{\varphi}_+}{t}
= 2 t \int_0^1 (\tilde{\psi}_A - \tilde{\psi}_V) u\,du\,,\\
&\tilde{\varphi}_+' + \tilde{\varphi}_-'
+ 2 i \bar{\Lambda} \tilde{\varphi}_+
= - 2 t \int_0^1 (\tilde{\psi}_A + \tilde{\psi}_X + 2 \tilde{\psi}_V u) du\,.
\end{split}
\label{eom5}
\end{equation}

The quark--antiquark--gluon distribution amplitudes in the momentum space
are defined as
\begin{equation}
\tilde{\psi}_i(t,u) = \int \psi_i(\omega,\xi)
e^{-i (\omega + \xi u) t} d\omega\,d\xi\,.
\label{psi2}
\end{equation}
Performing Fourier transform of the equations~(\ref{eom5}), we obtain
\begin{equation}
\begin{split}
&\omega \frac{d\varphi_-(\omega)}{d\omega} + \varphi_+(\omega) = I(\omega)\,,\\
&(\omega - 2 \bar{\Lambda}) \varphi_+(\omega) + \omega \varphi_-(\omega) = J(\omega)\,,
\end{split}
\label{eomm}
\end{equation}
where
\begin{align*}
&I(\omega) = 2 \frac{d}{d\omega}
\int_0^\omega d\rho \int_{\omega-\rho}^\infty \frac{d\xi}{\xi}
\frac{\partial}{\partial\xi}
\left[\psi_A(\rho,\xi) - \psi_V(\rho,\xi)\right]\,,\\
&J(\omega) = - 2 \frac{d}{d\omega}
\int_0^\omega d\rho \int_{\omega-\rho}^\infty \frac{d\xi}{\xi}
\left[\psi_A(\rho,\xi) + \psi_X(\rho,\xi)\right]
- 4 \int_0^\omega d\rho \int_{\omega-\rho}^\infty \frac{d\xi}{\xi}
\frac{\partial}{\partial\xi} \psi_V(\rho,\xi)\,.
\end{align*}
If we insist on having $\varphi_+(0)=0$, then the condition
\begin{equation}
J(0) = - 2 \int_0^\infty \frac{d\xi}{\xi}
\left[\psi_A(0,\xi) + \psi_X(0,\xi)\right] = 0
\label{cond0}
\end{equation}
must be satisfied.
It should follow from vanishing of $\psi_{A,X}(\omega,\xi)$ at $\omega\to0$
(behaviour of the three-particle distribution amplitudes
at the boundaries is discussed in~\cite{KMO:05}).

\begin{figure}[ht]
\begin{center}
\begin{picture}(42,45)
\put(21,24){\makebox(0,0){\includegraphics{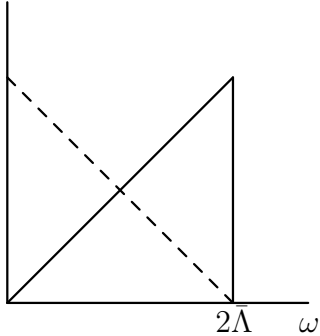}}}
\put(31,0){\makebox(0,0)[b]{$2\bar{\Lambda}$}}
\put(41,0){\makebox(0,0)[b]{$\omega$}}
\end{picture}
\end{center}
\caption{The Wandzura--Wilczek parts of $\varphi_+(\omega)$ (solid line)
and $\varphi_-(\omega)$ (dashed line)}
\label{F:WW}
\end{figure}

If we new $\psi_{A,V,X}(\omega,\xi)$,
we could solve the equations~(\ref{eomm}) for $\varphi_\pm(\omega)$.
This is not very realistic, because we don't know them.
The solution is a sum of two terms:
\begin{equation}
\varphi_\pm(\omega) = \varphi_\pm^{(WW)}(\omega) + \varphi_\pm^{(g)}(\omega)\,,
\label{ww}
\end{equation}
where $\varphi_\pm^{(WW)}(\omega)$ is the solution of~(\ref{eomm})
in the case if all quark--antiquark--gluon distribution amplitudes vanish
(it is called the Wandzura--Wilczek part of the solution),
and $\varphi_\pm^{(g)}(\omega)$ is induced by the gluonic terms.
The Wandzura--Wilczek part is
\begin{equation}
\varphi_+^{(WW)}(\omega)
= \frac{\omega}{2\bar{\Lambda}^2}
\theta(2\bar{\Lambda}-\omega)\,,\quad
\varphi_-^{(WW)}(\omega)
= \frac{2\bar{\Lambda}-\omega}{2\bar{\Lambda}^2}
\theta(2\bar{\Lambda}-\omega)
\label{WW}
\end{equation}
(Fig.~\ref{F:WW}); it satisfies the normalization conditions~(\ref{norm}).
Note that $2\bar{\Lambda}$ is the maximum value of $p_+$ of the light quark
if we assume that $B$-meson (having the residual energy $\bar{\Lambda}$)
consists of the on-shell heavy antiquark (always having zero residual energy)
and the on-shell light quark.
The gluon-induced part is given by some explicit integrals of $\psi_{A,V,X}(\omega,\xi)$;
we don't present these long expressions here, they can be found in~\cite{KKQT:01}.

The moments~(\ref{omn}) also consist of two contributions.
The Wandzura--Wilczek parts are
\begin{equation}
{<}\omega^n{>}_+^{(WW)}
= \frac{2(2\bar{\Lambda})^n}{n+2}\,,\quad
{<}\omega^n{>}_-^{(WW)}
= \frac{2(2\bar{\Lambda})^n}{(n+1)(n+2)}\,.
\label{WWmom}
\end{equation}
The gluon-induced part does not contribute to the zeroth moments (normalization)
and the first ones;
its contribution to the second moments~(\ref{mom2})
is expressed via the normalizations of the quark--antiquark--gluon
distribution amplitudes:
\begin{equation}
\begin{split}
&\int \psi_A(\omega,\xi) d\omega\,d\xi = \frac{1}{3} \lambda_E^2\,,\\
&\int \psi_V(\omega,\xi) d\omega\,d\xi = \frac{1}{3} \lambda_H^2\,,\\
&\int \psi_X(\omega,\xi) d\omega\,d\xi = 0\,.
\end{split}
\label{qqgnorm}
\end{equation}
The contributions to the higher moments are expressed via the moments
of $\psi_{A,V,X}(\omega,\xi)$; the explicit formulae can be found in~\cite{KKQT:01}.

\section{Evolution}
\label{S:Evol}

Until now, we neglected renormalization of the considered operators.
In this section we shall discuss renormalization of the leading-twist
$B$-meson distribution amplitude.

Let's summarize what we know about the bare operators of interest.
There are 3 families of such operators,
or 3 ``representations'' ($t$, $\omega$, $n$):
\begin{equation}
\begin{split}
&\tilde{O}_+(t) = \Q(0) \gamma_+ [0,z] q(z)\,,\\
&O_+(\omega) = \Q(0) \gamma_+ \delta(i D_+ - \omega) q(0)\,,\\
&O_+^{(n)} = \Q(0) \gamma_+ (i D_+)^n q(0)\,.
\end{split}
\label{bare}
\end{equation}
They can be converted into each other:
\begin{equation}
\begin{split}
&\tilde{O}_+(t)
= \int O_+(\omega) e^{-i \omega t} d\omega
= \sum_{n=0}^\infty O_+^{(n)} \frac{(-it)^n}{n!}\,,\\
&O_+(\omega)
= \int \tilde{O}_+(t) e^{i \omega t} \frac{dt}{2\pi}
= \int_{-i\infty}^{+i\infty} O_+^{(n)} \omega^{-n-1} \frac{dn}{2\pi i}\,,\\
&O_+^{(n)}
= \left.\left(i \frac{d}{dt}\right)^n O_+(t)\right|_{t=0}
= \int_0^\infty O_+(\omega) \omega^n d\omega\,.
\end{split}
\label{barerel}
\end{equation}
As we shall see, not all of these relations survive renormalization.

\begin{figure}[ht]
\begin{center}
\begin{picture}(100,23)
\put(16,10.5){\makebox(0,0){\includegraphics{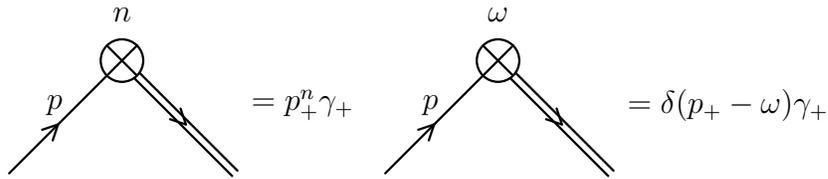}}}
\put(16,23){\makebox(0,0)[t]{$n$}}
\put(7,10){\makebox(0,0){$p$}}
\put(32,10){\makebox(0,0)[l]{${}=p_+^n\gamma_+$}}
\put(66,10.5){\makebox(0,0){\includegraphics{op1.eps}}}
\put(66,23){\makebox(0,0)[t]{$\omega$}}
\put(57,10){\makebox(0,0){$p$}}
\put(82,10){\makebox(0,0)[l]{${}=\delta(p_+-\omega)\gamma_+$}}
\end{picture}
\end{center}
\caption{Quark--antiquark vertices}
\label{F:Oqq}
\end{figure}

We shall calculate matrix elements of these bare operators,
therefore, we need the Feynman rules for them.
If we retain $i\partial_+$ in all brackets $(iD_+)^n$ in $O_+^{(n)}$,
we obtain the quark--antiquark vertex with $p_+^n$
shown in Fig.~\ref{F:Oqq}.
When transformed to the $\omega$-representation~(\ref{barerel}),
this gives $\delta(p_+-\omega)$,
as expected from the form~(\ref{bare}) of $O_+(\omega)$.

\begin{figure}[ht]
\begin{center}
\begin{picture}(100,25)
\put(16,12.5){\makebox(0,0){\includegraphics{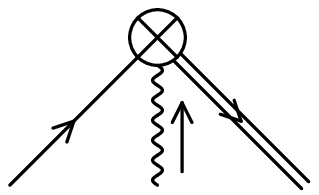}}}
\put(16,25){\makebox(0,0)[t]{$n$}}
\put(16,0){\makebox(0,0)[b]{$\mu\hspace{2mm}a$}}
\put(7,12){\makebox(0,0){$p$}}
\put(21,7){\makebox(0,0){$k$}}
\put(32,12){\makebox(0,0)[l]{$\displaystyle
{} = \frac{(p_++k_+)^n-p_+^n}{k_+} g_0 t^a n_+^\mu \gamma_+$}}
\end{picture}\\[5mm]
\begin{picture}(100,25)
\put(16,12.5){\makebox(0,0){\includegraphics{op2.eps}}}
\put(16,25){\makebox(0,0)[t]{$\omega$}}
\put(16,0){\makebox(0,0)[b]{$\mu\hspace{2mm}a$}}
\put(7,12){\makebox(0,0){$p$}}
\put(21,7){\makebox(0,0){$k$}}
\put(32,12){\makebox(0,0)[l]{$\displaystyle
{} = \frac{\delta(p_++k_+-\omega)-\delta(p_+-\omega)}{k_+}
g_0 t^a n_+^\mu \gamma_+$}}
\end{picture}
\end{center}
\caption{Quark--antiquark--gluon vertices}
\label{F:Oqqg}
\end{figure}

Now let's retain a single $A_+$ in $(iD_+)^n$.
After some combinatorics, this gives
\begin{equation}
(i \partial_+ + A_+)^n \to
\sum_{m=1}^n \left( \begin{array}{c} n \\ m \end{array} \right)
\left[(i\partial_+)^{m-1}A_+\right] (i \partial_+)^{n-m}\,.
\label{VQqg0}
\end{equation}
This gives the quark--antiquark--gluon vertex of $O_+^{(n)}$ with
\begin{equation}
\sum_{m=1}^n \left( \begin{array}{c} n \\ m \end{array} \right)
k_+^{m-1} p_+^{n-m} =
\frac{(p_++k_+)^n-p_+^n}{k_+}\,,
\label{VQqg}
\end{equation}
shown in Fig.~\ref{F:Oqqg}.
Transforming it to the $\omega$-representation~(\ref{barerel})
gives two $\delta$-functions (Fig.~\ref{F:Oqqg}).
Of course, the integral in $d\omega$ of this vertex vanishes,
because $O_+^{(0)}$ does not interact with gluons.

The bare operators $O_+(\omega)$ can be expressed
via the renormalized operators $O_+(\omega;\mu)$
($\mu$ is the $\overline{\text{MS}}$ renormalization scale),
or vice versa:
\begin{align}
O_+(\omega) &{}= \int Z_+(\omega,\omega';\mu) O_+(\omega';\mu) d\omega'\,,
\label{ren1}\\
O_+(\omega;\mu) &{}= \int Z_+^{-1}(\omega,\omega';\mu) O_+(\omega') d\omega'\,.
\label{ren2}
\end{align}
The renormalization ``matrices'' $Z_+(\omega,\omega';\mu)$
and $Z_+^{-1}(\omega,\omega';\mu)$ are inverse to each other:
\begin{equation}
\begin{split}
\int Z_+(\omega,\omega'';\mu) Z_+^{-1}(\omega'',\omega';\mu) d\omega''
&{}= \delta(\omega-\omega')\,,\\
\int Z_+^{-1}(\omega,\omega'';\mu) Z_+(\omega'',\omega';\mu) d\omega''
&{}= \delta(\omega-\omega')\,.
\end{split}
\label{inv}
\end{equation}

The renormalized operators $O_+(\omega;\mu)$
obey the renormalization-group equation
\begin{equation}
\frac{\partial O_+(\omega;\mu)}{\partial\log\mu}
+ \int \Gamma_+(\omega,\omega';\mu) O_+(\omega';\mu) d\omega' = 0\,,
\label{RG}
\end{equation}
where the anomalous dimension ``matrix'' is
\begin{equation}
\begin{split}
\Gamma_+(\omega,\omega';\mu) &{}=
\int Z_+^{-1}(\omega,\omega'';\mu)
\frac{\partial Z_+(\omega'',\omega';\mu)}{\partial\log\mu}
d\omega''\\
&{}= - \int
\frac{\partial Z_+^{-1}(\omega,\omega'';\mu)}{\partial\log\mu}
Z_+(\omega'',\omega';\mu) d\omega''\,.
\end{split}
\label{anomdim}
\end{equation}
This equation can be derived in two ways:
either we differentiate~(\ref{ren1}) in $d\log\mu$
and obtain 0 (because the bare operator is $\mu$-independent),
or we differentiate~(\ref{ren2}).

At one loop
\begin{equation}
\begin{split}
Z_+(\omega,\omega';\mu) &{}= \delta(\omega-\omega')
+ z_+^{(1)}(\omega,\omega';\mu) a_s
+ \cdots\\
Z_+^{-1}(\omega,\omega';\mu) &{}= \delta(\omega-\omega')
- z_+^{(1)}(\omega,\omega';\mu) a_s
+ \cdots\\
\Gamma_+(\omega,\omega';\mu) &{}=
\Gamma_+^{(1)}(\omega,\omega';\mu) a_s
+ \cdots
\end{split}
\label{Z1}
\end{equation}
where
\begin{equation*}
\displaystyle a_s=\frac{\alpha_s(\mu)}{4\pi}\,.
\end{equation*}
From~(\ref{anomdim}) we obtain
\begin{equation}
\Gamma_+^{(1)}(\omega,\omega';\mu)
= \frac{\partial z_+^{(1)}(\omega,\omega';\mu)}{\partial\log\mu}
- 2 \varepsilon z_+^{(1)}(\omega,\omega';\mu)\,.
\label{Gamma1}
\end{equation}

\begin{figure}[ht]
\begin{center}
\begin{picture}(106,42)
\put(16,31.5){\makebox(0,0){\includegraphics{op1.eps}}}
\put(16,10.5){\makebox(0,0){\includegraphics{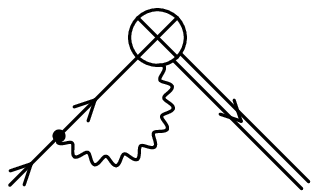}}}
\put(53,10.5){\makebox(0,0){\includegraphics{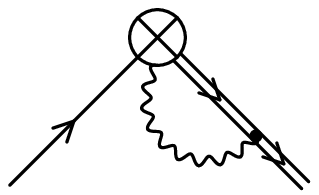}}}
\put(90,10.5){\makebox(0,0){\includegraphics{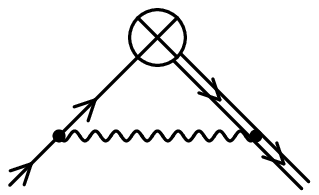}}}
\end{picture}
\end{center}
\caption{Diagrams for the matrix element of $O_+(\omega)$
between a quark--antiquark state and vacuum}
\label{F:M}
\end{figure}

The matrix element of $O_+(\omega)$
between a state with the light quark with momentum $p$
and the heavy antiquark with momentum $p'$
(which are off-shell) and vacuum is
\begin{equation}
\begin{split}
M ={}& {<}0|O_+(\omega)|q(p),\Q(p'){>}
= Z_q^{1/2} \tilde{Z}_Q^{1/2}
\left[ \delta(p_+-\omega) \gamma_+ + M_1 + M_2 + M_3 \right]\\
{}={}& {<}0|O_+(\omega;\mu)|q(p),\Q(p'){>}
+ a_s
\int z_+^{(1)}(\omega,\omega';\mu)
\delta(p_+-\omega') \gamma_+ d\omega'\,,
\end{split}
\label{M}
\end{equation}
where $M_{1,2,3}$ are the contributions of the one-loop diagrams
shown in Fig.~\ref{F:M},
the matrix element of the renormalized operator is finite
at $\varepsilon\to0$,
and $z_+^{(1)}(\omega,\omega';\mu)$ contains only
negative powers of $\varepsilon$
(in the term with $z_+^{(1)}$, we may substitute
$\delta(p_+-\omega')\gamma_+$ instead of ${<}0|O_+(\omega';\mu)|q(p),\Q(p'){>}$).
This allows us to find the renormalization ``matrix''
with one-loop accuracy.

\begin{figure}[ht]
\begin{center}
\begin{picture}(36,25)
\put(18,13.5){\makebox(0,0){\includegraphics{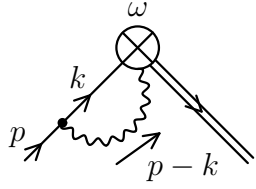}}}
\put(18,25){\makebox(0,0)[t]{$\omega$}}
\put(2,7){\makebox(0,0){$p$}}
\put(10,15){\makebox(0,0){$k$}}
\put(24,3){\makebox(0,0){$p-k$}}
\end{picture}
\end{center}
\caption{The first diagram}
\label{F:1}
\end{figure}

Let's calculate the first diagram (Fig.~\ref{F:1}) for
\begin{equation*}
p_+ = \omega'\,,\quad
p_\bot = 0\,,\quad
p^2 = p_ + p_- < 0\,.
\end{equation*}
It is
\begin{equation}
M_1 = i C_F g_0^2 \int \frac{d^d k}{(2\pi)^d}
\frac{\delta(p_+-\omega)-\delta(k_+-\omega)}{p_+-k_+}\,
\frac{\gamma_+ \rlap/k \gamma_+}%
{\left[-(p-k)^2-i0\right] \left[-k^2-i0\right]}\,.
\label{M1a}
\end{equation}
The numerator can be simplified as
\begin{equation*}
\gamma_+ \rlap/k \gamma_+ = 2 k_+ \gamma_+\,.
\end{equation*}

This diagram can be written in the form
\begin{equation}
M_1 = 2 C_F \frac{g_0^2 (-p^2)^{-\varepsilon}}{(4\pi)^{d/2}}
\Gamma(\varepsilon) \gamma_+
\left[ f_1(\omega,\omega')
- \delta(\omega-\omega') \int f_1(\omega'',\omega') d\omega'' \right]\,,
\label{M1b}
\end{equation}
where the term with $f_1(\omega,\omega')$
comes from the second $\delta$-function in~(\ref{M1a}).
This function is
\begin{equation*}
\pi^{d/2} (-p^2)^{-\varepsilon} \Gamma(\varepsilon) f_1(\omega,\omega')
= - i \frac{\omega}{\omega'-\omega} \int d^d k
\frac{\delta(k_+-\omega)}%
{\left[-(p-k)^2-i0\right] \left[-k^2-i0\right]}\,.
\end{equation*}
Now we use $\alpha$ parametrization
\begin{equation}
\frac{1}{-k^2-i0} = \int_0^\infty e^{(k^2+i0)\alpha} d\alpha
\label{alpha}
\end{equation}
for both denominators,
and also write the $\delta$-function as
\begin{equation}
\delta(k_+-\omega) = 2 \delta(2(k_+-\omega))
= 2 \int_{-\infty}^{+\infty} \exp\left[2(k_+-\omega)\nu\right]
\frac{d\nu}{2\pi}\,.
\label{delta}
\end{equation}
We obtain
\begin{equation*}
\begin{split}
&\pi^{d/2} (-p^2)^{-\varepsilon} \Gamma(\varepsilon) f_1(\omega,\omega')
= - 2 i \frac{\omega}{\omega'-\omega}
\int d^d k\,d\alpha_1\,d\alpha_2\,\frac{d\nu}{2\pi}\\
&{}\times\exp
\left[ \alpha_1 (k-p)^2 + \alpha_2 k^2 + 2 i \nu (k\cdot n_+ - \omega) \right]\,.
\end{split}
\end{equation*}
Shifting the integration momentum as
\begin{equation*}
k' = k - \frac{\alpha_1 p - i \nu n_+}{\alpha_1+\alpha_2}
\end{equation*}
to form a full square, we obtain
\begin{equation*}
\begin{split}
&- 2 i \frac{\omega}{\omega'-\omega}
\int d\alpha_1\,d\alpha_2
\exp\left[\frac{\alpha_1 \alpha_2}{\alpha_1+\alpha_2} p^2\right]\\
&\quad{}\times \int \frac{d\nu}{2\pi}
\exp\left[2 i \nu
\left( \frac{\alpha_1}{\alpha_1+\alpha_2} \omega' - \omega \right) \right]\,
\int d^d k' \exp\left[(\alpha_1+\alpha_2)(k^{\prime2}+i0)\right]
\end{split}
\end{equation*}
The integral in $d^d k'$ is calculated using the Wick rotation $k_0 = i k_{E0}$:
\begin{equation}
\int d^d k\,e^{\alpha (k^2+i0)}
= i \int d^d k_E\,e^{-\alpha k_E^2}
= i \left(\frac{\pi}{\alpha}\right)^{d/2}\,.
\label{Wick}
\end{equation}
The integral in $d\nu$ gives a $\delta$-function.
We obtain an integral in two $\alpha$-parameters:
\begin{equation*}
\begin{split}
&(-p^2)^{-\varepsilon} \Gamma(\varepsilon) f_1(\omega,\omega')\\
&{}= \frac{\omega}{\omega'-\omega}
\int d\alpha_1\,d\alpha_2\,(\alpha_1+\alpha_2)^{-d/2}
\exp\left[\frac{\alpha_1 \alpha_2}{\alpha_1+\alpha_2} p^2\right]\,
\delta\left(\frac{\alpha_1}{\alpha_1+\alpha_2}\omega'-\omega\right)\,.
\end{split}
\end{equation*}
It is always possible to calculate one integral in a ``radial'' variable $\eta$
in the space of $\alpha$-parameters via $\Gamma$-function.
In this case, the most convenient choice of such a variable
is $\eta=\alpha_1+\alpha_2$.
Therefore, we insert $\delta(\alpha_1+\alpha_2-\eta) d\eta$
under the integral sign, and substitute $\alpha_i = \eta x_i$:
\begin{equation*}
\frac{\omega}{\omega'-\omega}
\int d x_1\,d x_2\,\delta(x_1+x_2-1) \delta(x_1 \omega'-\omega)\,
\int d\eta \eta^{-1+\varepsilon} e^{- (-p^2) x_1 x_2 \eta}\,.
\end{equation*}
The final result is
\begin{equation}
f_1(\omega,\omega')
= \frac{\theta(\omega'-\omega)}{(\omega'-\omega)^{1+\varepsilon}}
\frac{\omega^{1-\varepsilon}}{(\omega')^{1-2\varepsilon}}\,.
\label{f1}
\end{equation}

Functions $F(\omega,\omega')$ which appear in the evolution kernel
should be understood as distributions: they are always integrated
with smooth test functions $\varphi(\omega')$.
The distribution $[F(\omega,\omega')]_+$ is defined by
\begin{equation}
\int \left[F(\omega,\omega')\right]_+ \varphi(\omega') d\omega'
= \int F(\omega,\omega') \left(\varphi(\omega')-\varphi(\omega)\right) d\omega'\,.
\label{plus}
\end{equation}
Therefore, formally we can write
\begin{equation}
F(\omega,\omega') = \left[F(\omega,\omega')\right]_+
+ \delta(\omega-\omega') \int F(\omega,\omega'') d\omega''\,.
\label{plus2}
\end{equation}
The result~(\ref{M1b}), (\ref{f1}) of the calculation
of the diagram in Fig.~\ref{F:1} can be written
via a $+$-distribution as
\begin{equation}
M_1 = 2 C_F \frac{g_0^2 (-p^2)^{-\varepsilon}}{(4\pi)^{d/2}}
\Gamma(\varepsilon) \gamma_+
\biggl[ \left[f_1(\omega,\omega')\right]_+
+ \delta(\omega-\omega')
\left( \int f_1(\omega,\omega'') d\omega''
- \int f_1(\omega'',\omega) d\omega'' \right) \biggr]\,.
\label{M1c}
\end{equation}
The coefficient of $\delta(\omega-\omega')$ here
can be calculated by substitution $x=\omega''/\omega$:
\begin{equation*}
\int_1^\infty x^{-1+2\varepsilon} (1-x)^{-1-\varepsilon} dx
- \int_0^1 x^{1-\varepsilon} (1-x)^{-1-\varepsilon} dx\,.
\end{equation*}
Substituting $x\to1/x$ in the first integral, we have
\begin{equation*}
\int_0^1 \left( x^{-\varepsilon} - x^{1-\varepsilon} \right)
(1-x)^{-1-\varepsilon} dx
= \int_0^1 x^{-\varepsilon} (1-x)^{-\varepsilon} dx\,.
\end{equation*}
Therefore, the final result for $M_1$ (Fig.~\ref{F:1}) is
\begin{equation}
M_1 = 2 C_F \frac{g_0^2 (-p^2)^{-\varepsilon}}{(4\pi)^{d/2}}
\Gamma(\varepsilon) \gamma_+
\left[ \left( \frac{\theta(\omega'-\omega)}{(\omega'-\omega)^{1+\varepsilon}}
\frac{\omega^{1-\varepsilon}}{(\omega')^{1-2\varepsilon}} \right)_+
+ \frac{\Gamma^2(1-\varepsilon)}{\Gamma(2-2\varepsilon)}
\delta(\omega-\omega') \right]\,.
\label{M1}
\end{equation}

\begin{figure}[ht]
\begin{center}
\begin{picture}(36,25)
\put(18,13.5){\makebox(0,0){\includegraphics{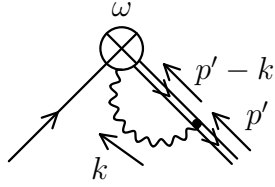}}}
\put(18,25){\makebox(0,0)[t]{$\omega$}}
\put(36,10){\makebox(0,0){$p'$}}
\put(33,16){\makebox(0,0){$p'-k$}}
\put(15,3){\makebox(0,0){$k$}}
\end{picture}
\end{center}
\caption{The second diagram}
\label{F:2}
\end{figure}

Now we shall calculate the second diagram (Fig.~\ref{F:2}) for
\begin{equation*}
p'\cdot v = \omega_1 < 0\,.
\end{equation*}
It is
\begin{equation}
M_2 = -i C_F g_0^2 \int \frac{d^d k}{(2\pi)^d}
\frac{\delta(p_++k_+-\omega)-\delta(p_+-\omega)}{k_+}\,
\frac{v_+ \gamma_+}{\left[-k^2-i0\right] \left[-(p'-k)\cdot v-i0\right]}\,.
\label{M2a}
\end{equation}

This diagram can be written in the form
\begin{equation}
M_2 = 2 C_F \frac{g_0^2}{(4\pi)^{d/2}}
\Gamma(\varepsilon) \gamma_+
\left[ f_2(\omega-\omega')
- \delta(\omega-\omega') \int f_2(\omega'') d\omega'' \right]\,,
\label{M2b}
\end{equation}
where the term with $f_2(\omega-\omega')$
comes from the first $\delta$-function in~(\ref{M2a}).
This function is
\begin{equation*}
\pi^{d/2} \Gamma(\varepsilon) f_2(\omega'')
= - \frac{i}{2\omega''} \int d^d k
\frac{\delta(k_+-\omega'')}%
{\left[-k^2-i0\right] \left[-(p'-k)\cdot v-i0\right]}\,.
\end{equation*}
Now we use $\alpha$-parametrization~(\ref{alpha}) for both denominators
(it is convenient to multiply the linear denominator by 2 before this)
and~(\ref{delta}) for the $\delta$-function, and obtain
\begin{equation*}
- \frac{2i}{\omega''}
\int d^d k\,d\alpha_1\,d\alpha_2\,\frac{d\nu}{2\pi}
\exp\left[ \alpha_1 k^2 + 2 \alpha_2 (p'-k) \cdot v
+ 2 i \nu (k\cdot n_+ - \omega'') \right]\,.
\end{equation*}
Shifting the integration momentum as
\begin{equation*}
k' = k - \frac{\alpha_2 v - i \nu n_+}{\alpha_1}
\end{equation*}
to form a full square, we obtain
\begin{equation*}
\begin{split}
&- \frac{2i}{\omega''}
\int d\alpha_1\,d\alpha_2
\exp\left[-\frac{\alpha_2^2}{\alpha_1} + 2 \omega_1 \alpha_2\right]\\
&\quad{}\times \int \frac{d\nu}{2\pi}
\exp\left[2 i \nu \left( \frac{\alpha_2}{\alpha_1} - \omega'' \right) \right]\,
\int d^d k' \exp\left[\alpha_1(k^{\prime2}+i0)\right]\,.
\end{split}
\end{equation*}
Taking the integrals in $d^d k'$ (\ref{Wick}) and $d\nu$,
we obtain the integral in two $\alpha$-parameters:
\begin{equation*}
\Gamma(\varepsilon) f_2(\omega'')
= \frac{1}{\omega''}
\int d\alpha_1\,d\alpha_2\,\alpha_1^{-d/2}
\exp\left[-\frac{\alpha_2^2}{\alpha_1} + 2 \omega_1 \alpha_2\right]\,
\delta\left(\frac{\alpha_2}{\alpha_1}-\omega''\right)\,.
\end{equation*}
Now the best choice of the ``radial'' variable is $\alpha_1$,
so we simply substitute $\alpha_2 = \alpha_1 y$:
\begin{equation*}
\frac{1}{\omega''} \int dy\,\delta(y-\omega'')
\int d\alpha_1 \alpha_1^{-1+\varepsilon}
e^{-y(y-2\omega_1)\alpha_1}\,.
\end{equation*}
Finally,
\begin{equation}
f_2(\omega'')
= \frac{\theta(\omega'')}{(\omega'')^{1+\varepsilon}
(\omega''-2\omega_1)^{\varepsilon}}\,.
\label{f2}
\end{equation}

Now we rewrite $M_2$ (\ref{M2b}), (\ref{f2}) via a $+$-distribution:
\begin{equation}
M_2 = 2 C_F \frac{g_0^2}{(4\pi)^{d/2}}
\Gamma(\varepsilon) \gamma_+
\biggl[ \left[f_2(\omega-\omega')\right]_+
+ \delta(\omega-\omega')
\left( \int_0^\omega f_2(\omega-\omega'') d\omega''
- \int_0^\infty f_2(\omega'') d\omega'' \right) \biggr]\,.
\label{M2c}
\end{equation}
The coefficient of $\delta(\omega-\omega')$ is
\begin{equation}
- \int_\omega^\infty f_2(\omega'') d\omega''\,.
\label{coefdelta}
\end{equation}
We introduced $\omega_1$ only to regularize possible infrared problems;
we only need the UV divergence of this diagram,
which does not depend on $\omega_1$.
Therefore, we may assume $|\omega_1|\ll\omega$.
The coefficient of $\delta(\omega-\omega')$ is
\begin{equation}
- \int_\omega^\infty (\omega'')^{-1-2\varepsilon} d\omega''
= - \frac{\omega^{-2\varepsilon}}{2\varepsilon}\,,
\label{coefdelta2}
\end{equation}
and the final result for $M_2$ (Fig.~\ref{F:2}) is
\begin{equation}
M_2 = 2 C_F \frac{g_0^2}{(4\pi)^{d/2}} \Gamma(\varepsilon) \gamma_+ \left[
\left(\frac{\theta(\omega-\omega')}{(\omega-\omega')^{1+2\varepsilon}}\right)_+
- \frac{\omega^{-2\varepsilon}}{2\varepsilon} \delta(\omega-\omega') \right]\,.
\label{M2}
\end{equation}

This one-loop diagram contains a $1/\varepsilon^2$ UV divergence!
It is in the coefficient of $\delta(\omega-\omega')$
given by the integral~(\ref{coefdelta}).
The function $f_2(\omega'')$ has a $1/\varepsilon$ UV divergence
in the integration in transverse momentum;
the longitudinal integral~(\ref{coefdelta})
is again UV divergent~(\ref{coefdelta2}).
The operator at the vertex of Fig.~\ref{F:2} is
(after integration in $\omega$) a light-like Wilson line;
the heavy-antiquark propagator is a time-like Wilson line.
The vertex correction to a time-like -- light-like cusp
on a Wilson line is known to have a $1/\varepsilon^2$
UV divergence at one loop~\cite{KK:92}.

\begin{figure}[ht]
\begin{center}
\begin{picture}(36,25)
\put(18,13.5){\makebox(0,0){\includegraphics{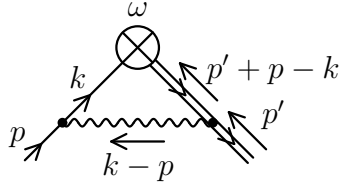}}}
\put(18,25){\makebox(0,0)[t]{$\omega$}}
\put(2,7){\makebox(0,0){$p$}}
\put(10,15){\makebox(0,0){$k$}}
\put(36,10){\makebox(0,0){$p'$}}
\put(36,16){\makebox(0,0){$p'+p-k$}}
\put(18,3){\makebox(0,0){$k-p$}}
\end{picture}
\end{center}
\caption{The third diagram}
\label{F:3}
\end{figure}

The third diagram (Fig.~\ref{F:3}) is
\begin{equation}
M_3 = -i C_F g_0^2 \int \frac{d^d k}{(2\pi)^d}
\frac{\delta(k_+-\omega)\,\gamma_+ \rlap/k \rlap/v}%
{\left[-(k-p)^2-i0\right] \left[-k^2-i0\right]
\left[-(p'+p-k)\cdot v-i0\right]}\,.
\label{M3a}
\end{equation}
The numerator can be simplified as
\begin{equation*}
\gamma_+ \rlap/k \rlap/v = k_+ \gamma_+\,.
\end{equation*}
We shall now demonstrate that this diagram is UV-finite
and hence does not contribute to the renormalization constant.

Using $\alpha$-parametrization~(\ref{alpha}) for the denominators
and~(\ref{delta}) for the $\delta$-function, and obtain
\begin{equation*}
\begin{split}
M_3 ={}& - 4 i C_F g_0^2 \omega \gamma_+
\int d^d k\,d\alpha_1\,d\alpha_2\,d\alpha_3\,\frac{d\nu}{2\pi}\\
&{}\times\exp\left[
\alpha_1 (k-p)^2 + \alpha_2 k^2 + 2 \alpha_3 (p+p'-k)\cdot v
+ 2 i \nu (k\cdot n_+ - \omega) \right]\,.
\end{split}
\end{equation*}
Shifting the integration momentum as
\begin{equation*}
k' = k - \frac{\alpha_1 p + \alpha_3 v - i \nu n_+}{\alpha_1 + \alpha_2}\,,
\end{equation*}
we obtain
\begin{equation*}
\begin{split}
M_3 ={}& - 4 i C_F g_0^2 \omega \gamma_+
\int d\alpha_1\,d\alpha_2\,d\alpha_3\,e^{-A}\\
&{}\times \int \frac{d\nu}{2\pi}
\exp\left[ 2 i \nu \left( \frac{\alpha_1\omega'+\alpha_3}{\alpha_1+\alpha_2}
- \omega \right) \right]\;
\int \frac{d^d k'}{(2\pi)^d} e^{(\alpha_1+\alpha_2)(k^{\prime2}+i0)}\\
{}={}& 2 C_F \frac{g_0^2}{(4\pi)^{d/2}} \omega \gamma_+
\int d\alpha_1\,d\alpha_2\,d\alpha_3\,
\delta\left(\frac{\alpha_1\omega'+\alpha_3}{\alpha_1+\alpha_2}-\omega\right)
e^{-A}\,,
\end{split}
\end{equation*}
where
\begin{equation*}
A = \frac{\alpha_3(\alpha_3-\alpha_2\omega')
+\alpha_2(\alpha_1\omega'+\alpha_3)(-p^2)}{\alpha_1+\alpha_2}
+\alpha_3(-2\omega)\,.
\end{equation*}
Inserting $\delta(\alpha_1+\alpha_2-\eta)d\eta$ under the integral sign
and making substitutions $\alpha_{1,2}=\eta x_{1,2}$, $\alpha_3=\eta y$,
we have $A=a\eta$, and the integral in the ``radial variable'' $\eta$
is easily calculated:
\begin{equation*}
\begin{split}
M_3 &{}= 2 C_F \frac{g_0^2}{(4\pi)^{d/2}} \omega \gamma_+
\int dx_1\,dy\,\delta(y+\omega'x_1-\omega)
\int d\eta\,\eta^{\varepsilon} e^{-a\eta}\\
&{}= 2 C_F \frac{g_0^2}{(4\pi)^{d/2}} \Gamma(1+\varepsilon)
\frac{\omega}{\omega'} \gamma_+
\int_{\max(0,\omega-\omega')}^\omega
\frac{dy}{a^{1+\varepsilon}}\,,
\end{split}
\end{equation*}
where
\begin{equation*}
a = \left[\omega - \omega' + \frac{\omega}{\omega^{\prime2}} (-p^2) - 2 \omega_1\right] y
+ (\omega - \omega') \frac{\omega}{\omega^{\prime2}} p^2\,.
\end{equation*}
The integral in $y$ is finite at $\varepsilon\to0$
(and easy to calculate).

Now we are ready to find the renormalization constant $Z_+(\omega,\omega';\mu)$.
Re-expressing
\begin{equation*}
\frac{g_0^2}{(4\pi)^{d/2}} =a_s \mu^{2\varepsilon} e^{\gamma_E\varepsilon}
\end{equation*}
in the matrix element~(\ref{M}), we have
\begin{equation}
\begin{split}
M ={}& \gamma_+ \left( Z_q \tilde{Z}_Q \right)^{1/2}
\biggl[ \delta(\omega-\omega') + 2 C_F \frac{\alpha_s(\mu)}{4\pi\varepsilon}\\
&{}\times\biggl(
\left(\frac{\theta(\omega'-\omega)}{\omega'-\omega}\frac{\omega}{\omega'}\right)_+
+ \delta(\omega-\omega')\\
&\hphantom{{}\times\biggl(\biggr.}
+ \left(\frac{\theta(\omega-\omega')}{\omega-\omega'}\right)_+
- \left( \frac{1}{2\varepsilon} - \log\frac{\omega}{\mu} \right)
\delta(\omega-\omega')
+ \mathcal{O}(\varepsilon) \biggr) \biggr]\,.
\end{split}
\label{Mren}
\end{equation}
The quark-field renormalization constants in QCD and HQET
th the Feynman gauge are
\begin{equation}
Z_q = 1 - C_F \frac{\alpha_s}{4\pi\varepsilon}\,,\quad
\tilde{Z}_Q = 1 + 2 C_F \frac{\alpha_s}{4\pi\varepsilon}\,.
\label{ZQq}
\end{equation}
Finally, the renormalization constant ``matrix'' at one loop is
\begin{equation}
\begin{split}
&Z_+(\omega,\omega';\mu) = \delta(\omega-\omega')
+ 2 C_F \frac{\alpha_s(\mu)}{4\pi\varepsilon}\\
&{}\times\biggl[\left(
\frac{\theta(\omega'-\omega)}{\omega'-\omega}\frac{\omega}{\omega'}
+\frac{\theta(\omega-\omega')}{\omega-\omega'}\right)_+
+ \left( - \frac{1}{2\varepsilon} + \log\frac{\omega}{\mu} + \frac{5}{4} \right)
\delta(\omega-\omega') \biggr]\,.
\end{split}
\label{Z1res}
\end{equation}

The evolution kernel~(\ref{anomdim}) has the structure
\begin{equation}
\Gamma_+(\omega,\omega';\mu) = \Gamma(\omega,\omega';a_s)
+ \left[ - \Gamma(\omega,\omega';a_s) \log\frac{\omega}{\mu}
+ \tilde{\gamma}_j(a_s) + \gamma(a_s) \right]\,.
\delta(\omega-\omega')\,.
\label{GammaStr}
\end{equation}
The logarithm $\log(\omega/\mu)$ only appears linearly,
to all orders of perturbation theory;
the coefficient of this logarithm
is the cusp anomalous dimension~\cite{KK:92}:
\begin{equation}
\Gamma(a_s) = \Gamma_0 a_s + \Gamma_1 a_s^2 + \cdots\quad
\Gamma_0 = 4 C_F
\label{Cusp}
\end{equation}
(the two-loop term is also known~\cite{KK:92}).
The non-logarithmic part is written as $\tilde{\gamma}_j+\gamma$,
where the anomalous dimension of the local current $j$~(\ref{j}) is
\begin{equation}
\tilde{\gamma}_j(a_s) = \tilde{\gamma}_{j0} a_s + \tilde{\gamma}_{j1} a_s^2
+ \cdots\quad
\tilde{\gamma}_{j0} = - 3 C_F
\label{gammaj}
\end{equation}
(the two- and three-loop terms are also known~\cite{BG:91,CG:03}).
It determines the evolution of $F(\mu)$;
the difference
\begin{equation}
\gamma(a_s) = \gamma_0 a_s + \gamma_1 a_s^2 + \cdots\quad
\gamma_0 = - 2 C_F
\label{gamma}
\end{equation}
appears in the evolution equation for $\varphi_+(\omega;\mu)$.
The non-$\delta$ term is
\begin{equation}
\Gamma(\omega,\omega',a_s) = \Gamma_0(\omega,\omega') a_s + \cdots\quad
\Gamma_0(\omega,\omega') = \left(
\frac{\theta(\omega'-\omega)}{\omega'-\omega}\frac{\omega}{\omega'}
+\frac{\theta(\omega-\omega')}{\omega-\omega'}\right)_+\,.
\label{Gamma}
\end{equation}
The evolution kernel has dimensionality of $1/\text{energy}$.

The evolution equation for the distribution amplitude is
\begin{equation}
\frac{\partial\varphi_+(\omega;\mu)}{\partial\log\mu}
+ \left[ - \Gamma(a_s) \log\frac{\omega}{\mu} + \gamma(a_s) \right]
\varphi_+(\omega;\mu)
+ \int \Gamma(\omega,\omega';a_s) \varphi_+(\omega';\mu) d\omega' = 0\,.
\label{evol}
\end{equation}
How to solve it?
Powers $\omega^n$ are eigenfunctions of the integral operator in~(\ref{evol}),
by dimensionality:
\begin{equation}
\int \Gamma(\omega,\omega';a_s) \omega^{\prime n} d\omega'
= \tilde{\Gamma}(n,a_s) \omega^n\,,
\label{eigen}
\end{equation}
where
\begin{equation}
\tilde{\Gamma}(n,a_s) = \tilde{\Gamma}_0 a_s + \cdots\quad
\tilde{\Gamma}_0(n) = 4 C_F \left[ \psi(1+n) + \psi(1-n) + 2 \gamma_E \right]\,,
\label{tildeGamma}
\end{equation}
from~(\ref{Gamma}) (here $\gamma_E$ is the Euler constant).
They are not eigenfunctions of the whole evolution operator,
due to the logarithmic term.
We can construct solutions with the power depending on $\mu$:
\begin{equation*}
\frac{\partial}{\partial\log\mu}
\left(\frac{\omega}{\mu_0}\right)^{n+\xi(\mu)}
= \left(\frac{\omega}{\mu_0}\right)^{n+\xi(\mu)}
\frac{d\xi}{d\log\mu} \log\frac{\omega}{\mu_0}\,.
\end{equation*}
If the function $\xi(\mu)$ obeys
\begin{equation}
\frac{d\xi}{d\log\mu} = \Gamma(a_s)\,,
\label{xi1}
\end{equation}
then $\log\omega$ will cancel in the evolution equation.

Dividing the definition~(\ref{xi1}) by
\begin{equation}
\frac{d\log a_s}{d\log\mu} = - 2 \beta(a_s)\,,\quad
\beta(a_s) = \beta_0 a_s + \beta_1 a_s^2 + \cdots
\label{beta}
\end{equation}
and integrating, we obtain
\begin{equation}
\xi = - \int_{a_{s0}}^{a_s} \frac{\Gamma(a_s)}{2\beta(a_s)} \frac{d a_s}{a_s}
= - \frac{\Gamma_0}{2\beta_0} \left[ \log\frac{a_s}{a_{s0}}
+ \left(\frac{\Gamma_1}{\Gamma_0} - \frac{\beta_1}{\beta_0}\right)
\left(a_s - a_{s0}\right) + \cdots \right]\,,
\label{xi}
\end{equation}
where $a_{s0}=\alpha_s(\mu_0)/(4\pi)$.
We shall use $\xi$ as an independent variable (``time'')
in the evolution equation instead of $\mu$.
The function
\begin{equation}
\varphi(\omega,\xi) = \left(\frac{\omega}{\mu_0}\right)^{n+\xi} e^{U(\xi)}
\label{solform}
\end{equation}
satisfies the evolution equation
\begin{equation}
\frac{\partial\log\varphi}{\partial\xi} - \log\frac{\omega}{\mu_0}
+ \frac{\tilde{\Gamma}(n+\xi,a_s)+\gamma(a_s)}{\Gamma(a_s)} = 0
\label{eqform}
\end{equation}
if
\begin{equation*}
\frac{dU}{d\xi} = - \log\frac{\mu}{\mu_0}
- \frac{\tilde{\Gamma}(n+\xi,a_s)+\gamma(a_s)}{\Gamma(a_s)}\,.
\end{equation*}
At the leading order,
\begin{equation*}
a_s = a_{s0} \exp\left(-\frac{2\beta_0}{\Gamma_0}\xi\right)\,,\quad
\log\frac{\mu}{\mu_0}
= \frac{2\pi}{\beta_0} \left(\frac{1}{a_s}-\frac{1}{a_{s0}}\right)
= \frac{2\pi}{\beta_0 a_{s0}}
\left[ \exp\left(\frac{2\beta_0}{\Gamma_0}\xi\right) - 1 \right]\,,
\end{equation*}
and
\begin{equation*}
U(\xi) = - \frac{2\pi}{\beta_0 a_{s0}} \left[ \frac{\Gamma_0}{2\beta_0}
\left(\exp\left(\frac{2\beta_0}{\Gamma_0}\xi\right) - 1\right) - \xi\right]
- \int_0^\xi \frac{\tilde{\Gamma_0}(n+\xi)}{\Gamma_0} d\xi
- \frac{\gamma_0}{\Gamma_0} \xi\,.
\end{equation*}
Here, from~(\ref{tildeGamma}),
\begin{equation*}
\int_0^\xi \frac{\tilde{\Gamma_0}(n+\xi)}{\Gamma_0} d\xi
= \log\frac{\Gamma(1+n+\xi)\Gamma(1-n)}{\Gamma(1-n-\xi)\Gamma(1+n)}
+ 2 \gamma_E \xi\,.
\end{equation*}
Finally, at the leading order,
\begin{equation}
e^{U(\xi)} =
\left(\frac{\Lambda_{\overline{\text{MS}}}}{\mu_0}\right)%
^{\frac{\Gamma_0}{2\beta_0}
\left[\exp\left(\frac{2\beta_0}{\Gamma_0}\xi\right)-1\right] - \xi}
\frac{\Gamma(1-n-\xi)\Gamma(1+n)}{\Gamma(1+n+\xi)\Gamma(1-n)}
\exp\left[-\left(\frac{\gamma_0}{\Gamma_0}+2\gamma_E\right)\xi\right]\,.
\label{Uxi}
\end{equation}

The distribution amplitude at $\mu_0$ can be expressed
via its moments~(\ref{intdn2}):
\begin{equation*}
\varphi_+(\omega;\mu_0) = \int_{-i\infty}^{+i\infty}
{<}\omega^{-1-n}{>}_+^{(\mu_0)} \omega^n \frac{dn}{2\pi i}\,.
\end{equation*}
Substituting the solutions~(\ref{solform})
instead of $\omega^n$ gives us the solution
of the evolution equation~(\ref{evol}) with initial conditions.
At the leading order~(\ref{Uxi}),
\begin{equation}
\begin{split}
\varphi_+(\omega;\mu) ={}&
\left(\frac{\Lambda_{\overline{\text{MS}}}}{\mu_0}\right)%
^{\frac{\Gamma_0}{2\beta_0}
\left[\exp\left(\frac{2\beta_0}{\Gamma_0}\xi\right)-1\right]}
\exp\left[-\left(\frac{\gamma_0}{\Gamma_0}+2\gamma_E\right)\xi\right]
\left(\frac{\omega}{\Lambda_{\overline{\text{MS}}}}\right)^\xi\\
&{}\times \int_{-\infty}^{+\infty}
{<}\omega^{-1-in}{>}_+^{(\mu_0)}\,\omega^{in}\,
\frac{\Gamma(1-n-\xi)\Gamma(1+n)}{\Gamma(1+n+\xi)\Gamma(1-n)}\,
\frac{dn}{2\pi}\,.
\end{split}
\label{sol}
\end{equation}

Qualitatively, each ``bin'' in the distribution amplitude at $\mu_0$
near some finite $\omega'$ produces, after evolution to a larger $\mu$,
a radiative tail, slowly decreasing (as $1/\omega$) at $\omega\gg\omega'$,
due to the evolution kernel $\Gamma(\omega,\omega';a_s)$
(which behaves as $1/\omega$ at large $\omega$, by dimensionality).
Therefore, the behaviour of the distribution amplitude
at large $\omega$ is $1/\omega$, up to logarithms.
The normalization integral~(\ref{norm}) of the distribution amplitude
logarithmically diverges at large $\omega$;
its moments~(\ref{omn}) with $n>0$ are power-divergent.
This means that the expression~(\ref{On})
for the local operators $O_+^{(n)}$ ($n\ge0$) via $O_+(\omega)$
is not valid for the renormalized operators.
Renormalization of $O_+(\omega)$ removes UV divergences
in transverse momenta;
in order to renormalize $O_+^{(n)}$,
we should also remove longitudinal UV divergences
(at large $\omega$).

\begin{figure}[ht]
\begin{center}
\begin{picture}(100,62)
\put(50,31){\makebox(0,0){\includegraphics[width=10cm]{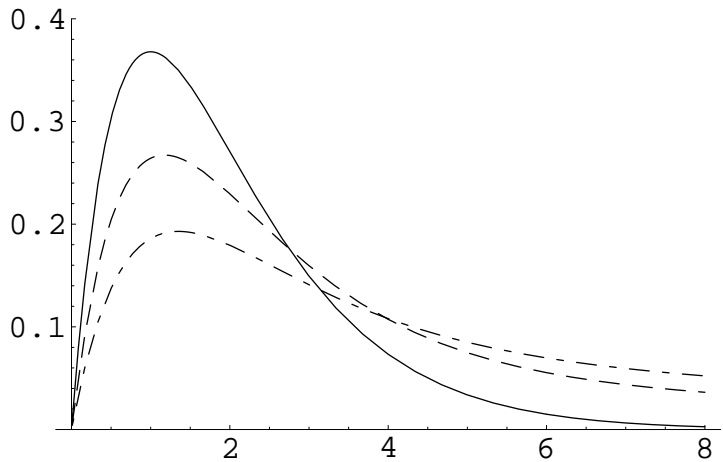}}}
\end{picture}
\end{center}
\caption{Evolution of $B$-meson distribution amplitude $\varphi_+(\omega;\mu)$}
\label{F:evol}
\end{figure}

As an illustration~\cite{LN:03}, let's suppose that at a low $\mu_0$
the distribution amplitude $\varphi_+(\omega)$
is given by the simple model~(\ref{model}).
Namely, this initial condition is taken at the scale $\mu_0$ where
\begin{equation*}
\alpha_s(\mu_0) = 1
\quad\Rightarrow\quad
\frac{\mu_0}{\Lambda_{\overline{\text{MS}}}}
= \exp\frac{2\pi}{\beta_0}\,,
\end{equation*}
where the quark model is supposed to work,
so that the light quark has only momenta
of order $\bar{\Lambda}$ (no radiative tail).
This function is shown by the solid line in Fig.~(\ref{F:evol}),
where $\omega$ is measured in units of $\omega_0$~(\ref{omega0}).
Using the leading-order solution~(\ref{sol})
of the evolution equation~(\ref{evol})
and supposing $\bar{\Lambda}/\Lambda_{\overline{\text{MS}}}=1.25$,
we obtain the distribution amplitude at the scale
where $\alpha_s(\mu)=0.5$ (dashed line)
and $0.3$ (dashed-dotted line).
We can see that the main part of the distribution amplitude
(at $\omega\sim\bar{\Lambda}$) becomes lower
and the radiative tail becomes more prominent
when $\mu$ grows.

\section{Sum rules}
\label{S:SR}

In order to estimate the quark--antiquark distribution amplitudes
$\varphi_\pm(\omega)$ of $B$-meson from QCD sum rules,
we consider the correlator of the local current $j_+$~(\ref{jP})
(having the quantum numbers of the ground-state meson)
and the bilocal operator $\tilde{O}_\pm(t)$~(\ref{Opmt}):
\begin{equation}
i{<}T \tilde{O}_\pm(t) \bar{\jmath}_+(-x) {>} =
\gamma_\pm \frac{1+\gamma^0}{2} \delta(\vec{x}\,) \theta(x^0)
\tilde{\Pi}_\pm(x^0,t)\,.
\label{sr1}
\end{equation}
We are most interested in its Fourier transform
\begin{equation}
\Pi_\pm(x^0,\omega) = \int \tilde{\Pi}_\pm(x^0,t) e^{i \omega t} \frac{dt}{2\pi}\,.
\label{sr2}
\end{equation}
Analytically continuing it from $x^0>0$ to $x^0=-i\tau$, we obtain
\begin{equation}
\Pi_\pm(\tau,\omega)
= \int \rho_\pm(\varepsilon,\omega) e^{-\varepsilon\tau} d\varepsilon
= F^2 \tilde{\varphi}_\pm(\omega) e^{-\bar{\Lambda}\tau}
+ \Pi_\pm^c(\tau,\omega)\,,
\label{sr3}
\end{equation}
where $\rho_\pm(\varepsilon,\omega)$ is the spectral density.
The correlator contains the contribution of the ground-state meson
(written explicitly in~(\ref{sr3}))
and of the continuum of excited states.

\begin{figure}[ht]
\begin{center}
\begin{picture}(32,17)
\put(16,9){\makebox(0,0){\includegraphics{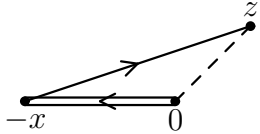}}}
\put(1,0){\makebox(0,0)[b]{$-x$}}
\put(21,0){\makebox(0,0)[b]{$0$}}
\put(31,17){\makebox(0,0)[t]{$z$}}
\end{picture}
\end{center}
\caption{Correlator of the local and the bilocal operators}
\label{F:SR}
\end{figure}

For sufficiently small $\tau$, we can calculate this correlator theoretically.
The perturbative contribution (Fig.~\ref{F:SR}) is given
in the fixed-point gauge $x^\mu A_\mu(x)=0$ by the light-quark propagator
from $-x$ to $z$:
\begin{equation}
\tilde{\Pi}_+^{(1)}(\tau,t) = \frac{N_c}{2 \pi^2 \tau (\tau+2it)^2}\,,\quad
\tilde{\Pi}_-^{(1)}(\tau,t) = \frac{N_c}{2 \pi^2 \tau^2 (\tau+2it)}\,.
\label{pert1}
\end{equation}
Its Fourier transform is
\begin{equation}
\Pi_+^{(1)}(\tau,\omega) = \frac{N_c}{8\pi^2\tau} \omega e^{-\omega\tau/2}\,,\quad
\Pi_-^{(1)}(\tau,\omega) = \frac{N_c}{4\pi^2\tau^2} e^{-\omega\tau/2}\,.
\label{pert2}
\end{equation}
Inverting the Laplace transform~(\ref{sr3}),
we find the spectral densities
(the integration contour should be to the right of the singularity at $\tau=0$):
\begin{equation}
\rho_\pm(\varepsilon,\omega) = \int_{-i\infty}^{+i\infty}
\Pi_\pm(\tau,\omega) e^{\varepsilon\tau} \frac{d\tau}{2\pi i}\,.
\label{pert3}
\end{equation}
We obtain
\begin{equation}
\rho_+^{(1)}(\varepsilon,\omega) = \frac{N_c}{8\pi^2} \omega\,
\theta\left(\varepsilon-\frac{\omega}{2}\right)\,,\quad
\rho_-^{(1)}(\varepsilon,\omega) = \frac{N_c}{4\pi^2}
\left(\varepsilon-\frac{\omega}{2}\right)\,
\theta\left(\varepsilon-\frac{\omega}{2}\right)\,.
\label{rho0}
\end{equation}
The perturbative spectral density $\rho_\pm^{(1)}(\varepsilon,\omega)$
describes the contribution of the on-shell quark--antiquark intermediate state
with energy $\varepsilon$ into the correlator~(\ref{sr3}).
The on-shell heavy antiquark has zero energy;
the maximum value of $p_+$ of the on-shell light quark with energy $\varepsilon$
is $2\varepsilon$.
This explains the $\theta$-functions in~(\ref{rho0}).

The quark condensate contribution is also important here.
It contains the vacuum average
\begin{equation*}
{<}\bar{q}(-x)[-x,0][0,z]q(z){>}\,.
\end{equation*}
This non-collinear quark condensate can be expanded
in terms of bilocal condensates~\cite{G:95},
the leading term in this expansion is the bilocal quark condensate
\begin{equation}
{<}\bar{q}(0)[0,x]q(x){>} = {<}\bar{q}q{>}\,f_S(x^2)
\label{bilocal}
\end{equation}
with $x\to x+z$.
This leading term produces the same contribution into $\tilde{\Pi}_\pm$:
\begin{equation}
\tilde{\Pi}_\pm^{(2)}(x^0,t) = - \frac{1}{4} {<}\bar{q}q{>} f_S((x+z)^2)\,.
\label{cond1}
\end{equation}
The expansion of the bilocal quark condensate at small $x$ is
\begin{equation}
f_S(x^2) = 1 + \frac{m_0^2}{16} x^2 + \cdots
\label{local}
\end{equation}
where
\begin{equation}
{<}\bar{q} G_{\mu\nu} \sigma^{\mu\nu} q{>} = m_0^2 {<}\bar{q}q{>}\,.
\label{m02}
\end{equation}
In general, it can be written as
\begin{equation}
f_S(x^2) = \int \tilde{f}_S(\nu) e^{\nu x^2} d\nu\,,
\label{tildefS}
\end{equation}
where $\tilde{f}_S(\nu)$ has the meaning
of the virtuality distribution function of vacuum quarks.
Its moments are expressed via vacuum averages of local operators:
\begin{equation}
\int \tilde{f}_S(\nu) d\nu = 1\,,\quad
\int \tilde{f}_S(\nu) \nu d\nu = \frac{m_0^2}{16}\,,\quad\ldots
\label{momfS}
\end{equation}
In terms of this function, the quark-condensate contribution
into the Fourier-transformed correlators is
\begin{equation}
\Pi_\pm^{(2)}(\tau,\omega) = - \frac{{<}\bar{q}q{>}}{8\tau}
\tilde{f}_S\left(\frac{\omega}{2\tau}\right) e^{-\omega\tau/2}\,.
\label{cond2}
\end{equation}
In other words, the virtuality distribution function
appears directly in the sum rules for the $B$-meson distribution amplitudes!
This is similar to the case of non-diagonal sum rules
for the pion distribution amplitude~\cite{BM:95}.

The local operator expansion~(\ref{local}) gives
\begin{equation}
\tilde{f}_S(\nu) = \delta(\nu) - \frac{m_0^2}{16} \delta'(\nu) + \cdots
\label{localfS}
\end{equation}
Of course, a smooth function can always be expanded in derivatives
of $\delta$-function, but such an expansion does not tell us much
about the shape of this function,
unless we sum an infinite number of terms.
The bilocal quark condensate~(\ref{bilocal}) at large $-x^2$ behaves as
\begin{equation}
f_S(x^2) \sim e^{-\bar{\Lambda}\sqrt{-x^2}}\,,
\label{asympt}
\end{equation}
because the Wilson line $[0,x]$ can be considered an HQET heavy-quark propagator,
and this vacuum average is the correlator of two HQET heavy--light currents,
having $B$-meson as the lowest-energy intermediate state.
Of course, the conditions~(\ref{momfS}), (\ref{asympt})
don't determine the shape of the distribution function $\tilde{f}_S(\nu)$
in a unique way.
They are satisfied by the Bakulev--Mikhailov ansatz~\cite{BM:95}
\begin{equation}
\tilde{f}_S(\nu) = N \exp\left(-\frac{\bar{\Lambda}^2}{4\nu}-\sigma\nu\right)\,,
\label{ansatz}
\end{equation}
where the parameters $\sigma$, $N$ are determined by~(\ref{momfS}).

Now we equate the theoretical result for the correlators
to the result obtained from a phenomenological model
of the spectral densities.
They have the contribution $\sim\delta(\varepsilon-\bar{\Lambda})$
of the ground-state meson and that of the continuum of excited states.
As usual, this contribution is modeled
by the perturbative spectral densities~(\ref{rho0})
starting from a continuum threshold energy $\varepsilon_c$:
\begin{equation}
\rho_\pm(\varepsilon,\omega)
= F^2 \varphi_\pm(\omega) \delta\left(\varepsilon-\bar{\Lambda}\right)
+ \rho_\pm^{(1)}(\varepsilon,\omega) \theta(\varepsilon-\varepsilon_c)\,.
\label{rhophen}
\end{equation}
With this model, the equality of the phenomenological expression
for the correlator and the theoretical one,
\begin{equation*}
F^2 \varphi_\pm(\omega) e^{-\bar{\Lambda}\tau}
+ \int_{\varepsilon_c}^\infty \rho_\pm^{(1)}(\varepsilon,\omega)
e^{-\varepsilon\tau} d\varepsilon
= \int_0^\infty \rho_\pm^{(1)}(\varepsilon,\omega)
e^{-\varepsilon\tau} d\varepsilon
+ \Pi_\pm^{(2)}(\tau,\omega)\,,
\end{equation*}
becomes
\begin{equation}
F^2 \varphi_\pm(\omega) e^{-\bar{\Lambda}\tau}
= \int_0^{\varepsilon_c} \rho_\pm^{(1)}(\varepsilon,\omega)
e^{-\varepsilon\tau} d\varepsilon
+ \Pi_\pm^{(2)}(\tau,\omega)\,,
\label{sr}
\end{equation}
where the perturbative spectral density is integrated
over the ``duality interval'' of the ground-state meson
(from $0$ to the continuum threshold $\varepsilon_c$).
We obtain the sum rules for the $B$-meson distribution amplitudes
\begin{equation}
\begin{split}
&F^2 \varphi_+(\omega) e^{-\bar{\Lambda}\tau}
= \frac{N_c}{8\pi^2\tau} \omega e^{-\omega\tau/2}
\delta_0\left(\left(\varepsilon_c-\frac{\omega}{2}\right)\tau\right)
- \frac{{<}\bar{q}q{>}}{8\tau}
\tilde{f}_S\left(\frac{\omega}{2\tau}\right) e^{-\omega\tau/2}\,,\\
&F^2 \varphi_-(\omega) e^{-\bar{\Lambda}\tau}
= \frac{N_c}{4\pi^2\tau^2} e^{-\omega\tau/2}
\delta_1\left(\left(\varepsilon_c-\frac{\omega}{2}\right)\tau\right)
- \frac{{<}\bar{q}q{>}}{8\tau}
\tilde{f}_S\left(\frac{\omega}{2\tau}\right) e^{-\omega\tau/2}\,.
\end{split}
\label{SR}
\end{equation}
where the functions
\begin{equation}
\delta_n(x) = \theta(x) \left(1 - e^{-x} \sum_{m=0}^n \frac{x^m}{m!}\right)
\label{deltan}
\end{equation}
describe the effect of subtracting continuum from the perturbative contribution
when the spectral density is $\sim\varepsilon^n$.
The perturbative contributions vanish at $\omega>2\varepsilon_c$,
because of the properties of the spectral densities~(\ref{rho0}).

\begin{figure}[p]
\begin{center}
\begin{picture}(100,62)
\put(50,31){\makebox(0,0){\includegraphics[width=10cm]{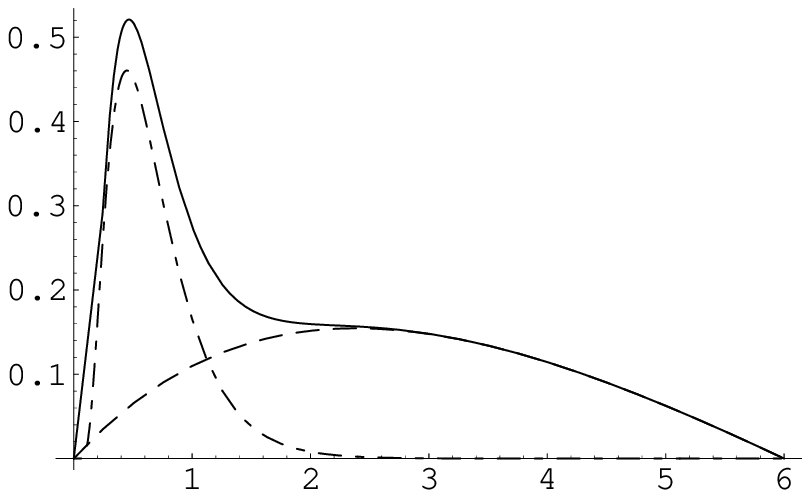}}}
\end{picture}
\end{center}
\caption{Distribution amplitude $\varphi_+(\omega)$ (solid line),
perturbative contribution (dashed line),
and quark-condensate contribution (dashed-dotted line)}
\label{F:p}
\end{figure}

\begin{figure}[p]
\begin{center}
\begin{picture}(100,62)
\put(50,31){\makebox(0,0){\includegraphics[width=10cm]{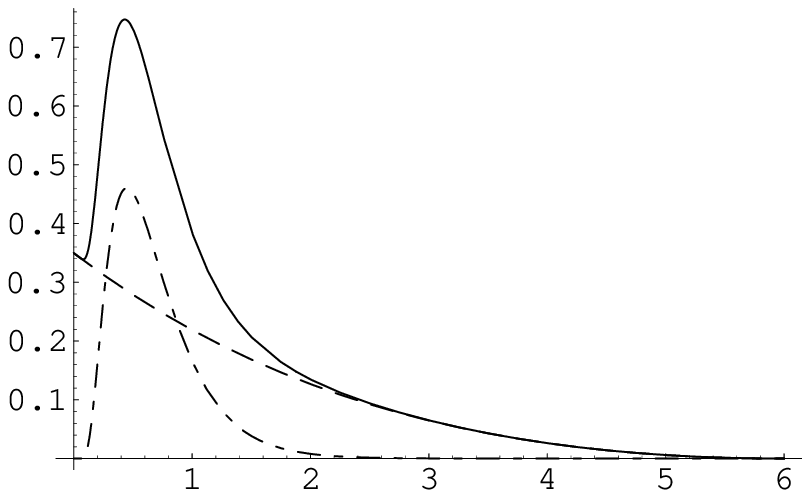}}}
\end{picture}
\end{center}
\caption{Distribution amplitude $\varphi_-(\omega)$ (solid line),
perturbative contribution (dashed line),
and quark-condensate contribution (dashed-dotted line)}
\label{F:m}
\end{figure}

Setting $t=0$ in the correlator~(\ref{sr1}),
or integrating~(\ref{sr2}) in $d\omega$,
we obtain the well-known sum rule for $F^2$:
\begin{equation}
F^2 e^{-\bar{\Lambda}\tau}
= \frac{N_c}{2\pi^2\tau^3} \delta_2(\varepsilon_c\tau)
- \frac{1}{4} {<}\bar{q}q{>} f_S(-\tau^2)\,.
\label{SR0}
\end{equation}
The natural energy scale in this sum rule is
\begin{equation}
k = \left( - \frac{\pi^2}{2 N_c} {<}\bar{q}{q} \right)^{1/3}
\approx 260\,\text{MeV}\,.
\label{k}
\end{equation}
With $m_0/(4k)=0.85$, one finds the optimal value
of the continuum threshold $\varepsilon_c/k=3$;
then a wide plato at $1/(k\tau)\in[1.7,2.5]$ exists,
and yields $\bar{\Lambda}/k=1.65$.
We divide the sum rules~(\ref{SR}) by~(\ref{SR0})
and take $1/(k\tau)=2$ (in the middle of the plato).
The results are shown in Fig.~\ref{F:p}, \ref{F:m},
where $\omega$ is measured in units of $k$.
They are automatically normalized~(\ref{norm}).
Of course, the leading-order sum rules cannot tell us
at what scale these distribution amplitudes are normalized;
this scale must be low, $\mu\sim1/\tau$.
The perturbative contributions vanish at $\omega>2\varepsilon_c$.
The quark-condensate contribution is the same
for both $\varphi_+(\omega)$ and $\varphi_-(\omega)$.
Here we used the ansatz~(\ref{ansatz}).
This contribution gives a sharp peak at low $\omega$%
\footnote{For the pion distribution amplitude,
the quark-condensate contribution gives enhancements
near $x=0$ and $1$.}.
Details of its shape are unknown,
but it cannot be wider because of the restriction
on the first moment~(\ref{momfS}).
This contribution falls off quickly at larger $\omega$.
Therefore, the distribution amplitudes at this low $\mu$
are only non-zero at $\omega\sim\bar{\Lambda}$,
in accord with the expectations of the quark model.

\begin{figure}[ht]
\begin{center}
\begin{picture}(42,36)
\put(21,20){\makebox(0,0){\includegraphics{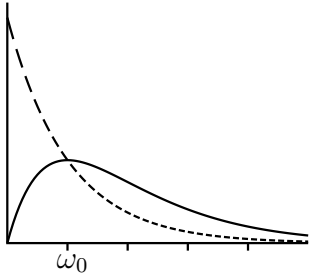}}}
\put(9.5,0){\makebox(0,0)[b]{$\omega_0$}}
\end{picture}
\end{center}
\caption{Model distribution amplitudes: $\varphi_+(\omega)$ (solid line)
and $\varphi_-(\omega)$ (dashed line)}
\label{F:model}
\end{figure}

If we neglect the quark-condensate contribution
(though this is not a good idea)
and also the effect of continuum subtraction%
\footnote{If the same procedure is applied to the diagonal sum rule
for the pion distribution amplitude, it yields the asymptotic shape.},
then the perturbative contributions~(\ref{pert2})
suggest the following simple model of distribution amplitudes:
\begin{equation}
\varphi_+(\omega) = \frac{\omega}{\omega_0^2} e^{-\omega/\omega_0}\,,\quad
\varphi_-(\omega) = \frac{1}{\omega_0} e^{-\omega/\omega_0}\,.
\label{model}
\end{equation}
They are normalized~(\ref{norm});
from the first moments~(\ref{mom1}) we obtain
\begin{equation}
\omega_0 = \frac{2}{3} \bar{\Lambda}\,.
\label{omega0}
\end{equation}
These functions (Fig.~\ref{F:model}) have something in common
with the Wandzura--Wilczek ones (Fig.~\ref{F:WW}),
but in contrast to them they fall off smoothly at large $\omega$.

Radiative corrections to the perturbative spectral density
and the dimension-3 quark-condensate contribution for $\varphi_+$
were calculated in~\cite{BIK:04}.
With these corrections, one can check that the correlator~(\ref{sr2})
satisfies the evolution equation~(\ref{RG}).
The resulting sum rules should be used together with the sum rules
for $F^2$ with radiative corrections~\cite{BG:92}.
The perturbative spectral density is
\begin{equation}
\begin{split}
&\rho_+^{(1)}(\varepsilon,\omega) = \frac{N_c}{8\pi^2} \omega\\
&{}\times\left\{
\begin{array}{ll}
x>1:&\displaystyle
1 + C_F \frac{\alpha_s}{4\pi} \biggl[
- 2 \log^2 \frac{\omega}{\mu} - 4 \bigl(\log(x-1)+1\bigr) \log\frac{\omega}{\mu}\\
&\displaystyle{}
+ 2 \Li2\left(\frac{1}{1-x}\right) - \log^2 (x-1)\\
&\displaystyle{}
- (2x+3) \log(x-1) + 2 x \log x + \frac{7}{12} \pi^2 + 7 \biggr]\\
x<1:&\displaystyle
2 C_F \frac{\alpha_s}{4\pi} \biggl[
2 \bigl(\log(1-x)+x\bigr) \log\frac{\omega}{\mu} + 2 \log^2 (1-x)\\
&\displaystyle{}
+ (2x-1) \log(1-x) - x \biggr]
\end{array}
\right.
\end{split}
\label{rho1}
\end{equation}
where $x = 2\varepsilon/\omega$.
Now the spectral density does not vanish at $\varepsilon<\omega/2$,
it is just suppressed by $\alpha_s/(4\pi)$.
Therefore, the perturbative contribution
to the sum rule~(\ref{SR}) for $\varphi_+(\omega)$
does not vanish at $\omega>2\varepsilon_c$.
It produces a radiative tail $\sim1/\omega$
with the magnitude of order $\alpha_s/(4\pi)$.

The radiative correction to the dimension-3 quark-condensate contribution
cannot be used together with a model of bilocal condensate,
because radiative corrections to higher-dimensional contributions
are not known.
It is vital to use some model of the bilocal quark condensate
in the sum rules for the distribution amplitude,
because the local operator expansion produces contributions
$\delta(\omega)$, $\delta'(\omega)$, \dots, which don't tell us much
about the shape of the distribution amplitude.
Therefore, the full result for this correction
cannot be used in the sum rule for $\varphi_+(\omega)$.
Fortunately, the authors of~\cite{BIK:04} demonstrated
that a part of this correction is universal,
and exponentialized into the Sudakov factor.
We may multiply the bilocal quark-condensate contribution
by this Sudakov factor.

\section*{Acknowledgments}

I am grateful to M.~Neubert for collaboration on~\cite{GN:97}
and to S.V.~Mikhailov for useful discussions.

\end{document}